%% file: HosseinigokiKosut_IEEEIT_Dec2017.tex
\documentclass[draftclsnofoot, 12pt, onecolumn]{IEEEtran} 

\input{header.tex}
\let\labelindent\relax
\usepackage{enumitem}
\usepackage{bbm}
\usepackage{cite}
\usepackage[usenames, dvipsnames]{color}
\usepackage{soul}
\usepackage{hyperref}
 
\usepackage{subfig}
\usepackage{graphicx}

\allowdisplaybreaks

\IEEEoverridecommandlockouts

\title{\LARGE \bf
The Gaussian Interference Channel in the Presence of Malicious Jammers}

\author{Fatemeh Hosseinigoki$^{1}$ and Oliver Kosut$^{1}$
\thanks{This material is based upon work supported by the National Science Foundation under Grant No. CCF-1453718.}
\thanks{This paper was presented in part at the 54th Annual Allerton Conference on Communication, Control, and Computing, Allerton 2016 \cite{AllertonPaper}.}
\thanks{$^{1}$School of Electrical, Computer and Energy Engineering, Arizona State University, Tempe, AZ 85287.
        {\tt\small \{fhossei1,okosut\}@asu.edu}}}

\begin{document}
\maketitle

\newcommand{\T}{\calT_{\eps}^{(n)}}

\begin{abstract}
This paper considers the two-user Gaussian interference channel in the presence of adversarial jammers. We first provide a general model including an arbitrary number of jammers, and show that its capacity region is equivalent to that of a simplified model in which the received jamming signal at each decoder is independent. Next, existing outer and inner bounds for two-user Gaussian interference channel are generalized for this simplified jamming model. We show that for certain problem parameters, precisely the same bounds hold, but with the noise variance increased by the received power of the jammer at each receiver. Thus, the jammers can do no better than to transmit Gaussian noise. For these problem parameters, this allows us to recover the half-bit theorem. In weak and strong interference regime, our inner bound matches the corresponding Han-Kobayashi bound with increased noise variance by the received power of the jammer, and even in strong interference we achieve the exact capacity. Furthermore, we determine the symmetric degrees of freedom where the signal-to-noise, interference-to-noise and jammer-to-noise ratios are all tend to infinity. Moreover, we show that, if the jammer has greater received power than the legitimate user, symmetrizability makes the capacity zero. The proof of the outer bound is straightforward, while the inner bound generalizes the Han-Kobayashi rate splitting scheme. As a novel aspect, the inner bound takes advantage of the common message acting as common randomness for the private message; hence, the jammer cannot symmetrize only the private codeword without being detected. This complication requires an extra condition on the signal power, so that in general our inner bound is not identical to the Han-Kobayashi bound. We also prove a new variation of the packing lemma that applies for multiple Gaussian codebooks in an adversarial setting.
\end{abstract}

\makeatletter
\newcommand{\BBigg}{\bBigg@{3.0}}
\newcommand{\vast}{\bBigg@{3.5}}
\newcommand{\vastt}{\bBigg@{4.0}}
\newcommand{\Vast}{\bBigg@{4.5}}
\makeatother

\section{Introduction}

The open nature of the wireless communication medium makes it inherently vulnerable to an active attack, wherein a malicious adversary (or jammer) transmits into the medium to disrupt the operation of the legitimate users. Therefore, developing techniques to manage the presence of an active attacker and to characterize the effect of an attacker on the fundamental limits of wireless communication is important. In this paper, we investigate the two-user Gaussian interference channel (GIC) in the presence of intelligent jammers (see Fig.~\ref{fig:Fig1Label}) to characterize the capacity region under the input and state constraints.

\subsection{Prior Work}

There are a wide variety of information-theoretic studies of active attacks in the literature such as the arbitrarily-varying channel (AVC) \cite{Ahlswede1978,Csiszar1988,Csiszar}, correlated jamming \cite{Medard1997}, network coding with adversarial errors \cite{Yeung2006,Cai2006}, multiuser channels with correlated jamming \cite{Shabnam2009}, and a diamond network in the presence of a jammer \cite{Mohajer2010}. 

In \cite{Ahlswede1978}, Ahlswede proved that the capacity of discrete memoryless AVC for deterministic codes is either zero or equal to the random code capacity under the average probability of error criterion. Ericson in \cite{Ericson} found a sufficient condition for the capacity to be zero, and later in \cite{Csiszar1988} it is proved that the condition is also necessary. Csiszar and Narayan further studied the deterministic code capacity of the Gaussian AVC under the average probability of error criterion in \cite{Csiszar} while they also considered the input and state power constraints. Here, we mainly follow the Gaussian AVC of \cite{Csiszar} in which an adversary does not have any knowledge about the legitimate user's signal (but does know the code), and it may send an arbitrary sequence across the coding block subject to a power constraint. 

On the other hand, the interference channel (without a jammer) is one of the fundamental problems in network information theory, and the exact capacity region is still unknown in general. However, the Han-Kobayashi inner bound \cite{Han1981} is optimal or near-optimal for many interference channels. The proof of this inner bound involves rate splitting, wherein each transmitter sends a common message, decoded by both receivers, as well as a private message, decoded by only the intended receiver. The authors in \cite{Etkin2008} showed that for the GIC, the Han-Kobayashi comes within half a bit of the capacity region. Furthermore, \cite{Annapureddy} obtains an outer bound on the capacity region of the GIC, and it is shown that for sufficiently weak interference signals, treating interference as noise achieves the sum capacity. The deterministic interference channel model is proposed by Brestler and Tse \cite{Bresler}, and they show that the capacity of this channel is within a constant number of bits of the corresponding GIC.

\subsection{Main Results}

Our main contribution is to generalize existing inner and outer bounds for the GIC in the presence of AVC-style jammers. We provide a generalized GIC model with $G$ jammers ($G\geq 1$), and show that the capacity region is equivalent to the capacity region of GIC with only two independent jammers. We show that the capacity region depends only on the received power of the jamming signal, not on the number of jammers. Moreover, we obtain the symmetric degrees of freedom (DoF) by taking the limit of the normalized symmetric capacity as signal-to-noise, interference-to-noise and jammer-to-noise ratios converge to infinity. This characterization generalized the so-called ``W'' DoF curve in \cite[p. 153]{ElGamal}. We also recover the optimal sum-rate for the weak interference regime, as well as the exact capacity region for the strong interference regime. We show that our bounds are within a half-bit in some regions, but we cannot show that this holds in general.

We show that the outer bound in \cite{Etkin2008} holds with the noise variance increased by the received power of the corresponding jammer at each receiver. The proof, given in Section~\ref{4}, follows by applying the outer bound in \cite{Etkin2008} with the jammers choosing to transmit Gaussian noise. Moreover, we show that if the jammer's received power at either receiver is larger than that of the intended transmitter, AVC symmetrizability prevents this message from being decoded, because the receiver cannot distinguish the legitimate codeword from the jammer's counterfeit; thus the capacity becomes zero.

We also provide a generalization of the Han-Kobayashi inner bound. For certain problem parameters---for example, in the symmetric case when the jammer's received power is less than that of the interfering user---this inner bound is precisely the Han-Kobayashi inner bound with the noise variance again increased by the received power of the jammer. Thus, for these problem parameters we recover the half-bit theorem of \cite{Etkin2008}, although we cannot prove that it holds in general. The proof of the inner bound, given in Section~\ref{5}, is somewhat more involved, as the receivers must decode correctly no matter what the jammer transmits. One novel aspect of our inner bound proof is that we use the common message in the rate-splitting scheme as common randomness for the private message. Thus, if the jammer has more power than the private codeword but less than both together, it cannot use symmetrization without being detected, and thus the receiver can decode. 

In both \cite{Csiszar1988} and \cite{Csiszar}, Csiszar and Narayan utilized lemmas (Lemma 3 in \cite{Csiszar1988} and Lemma 1 in \cite{Csiszar}) which assert the existence of codebooks with desirable properties in order to prove achievability results. However, they only provided these lemmas for a single codebook, and for either discrete random vectors (in \cite{Csiszar1988}) or codewords uniformly distributed on the unit ball (in \cite{Csiszar}). On the other hand, our proof requires a variation on the Gaussian AVC packing lemma that handles decoding of multiple superposed Gaussian codebooks. Our main technical tools for this goal are Lemma~\ref{lem4}, Lemma~\ref{lem3} and Lemma~\ref{lem5}, proved in Appendix \ref{ApnB}, \ref{ApnA} and \ref{ApnC}, respectively.

The remainder of this paper is organized as follows. We describe the problem and the system model for GIC with jammers in Section \ref{2}. Our main results including the outer and the inner bound for the capacity region are given by two theorems in Section \ref{3}. We also discuss implications of our bounds for different regimes, as well as illustrate numerical results in Section \ref{4}. The proof for the outer bound is provided in Section \ref{5}, and Section \ref{6} consists of the inner bound proof. We conclude in Section \ref{7}.

\emph{Notations:} We use bold letters to indicate the $n$-length vectors. Notation $\mathbbm{1}(\cdot)$ refers to the indicator function. We employ $\langle \cdot,\cdot \rangle $ and $\|\cdot\|$ to denote inner product and norm-2, respectively. We indicate the positive-part function and the expectation by $|\cdot|^+$ and $\mathbb{E}[\cdot]$, respectively. Also, for an integer $N$, notation $[N]$ stands for the set $\{1,2,3,\ldots,N\}$. Notation $\mathbf{I}_n$ represents the identity matrix of size $n$. Each of $\log(\cdot)$ and $\exp(\cdot)$ functions has base 2. Moreover, $C(x)=\frac{1}{2}\log(1+x)$ and $\bar{\alpha} = 1-\alpha$.

\section{Problem Statement} \label{2}

\begin{figure}[t]
\centering
    \includegraphics[width=0.5\linewidth]{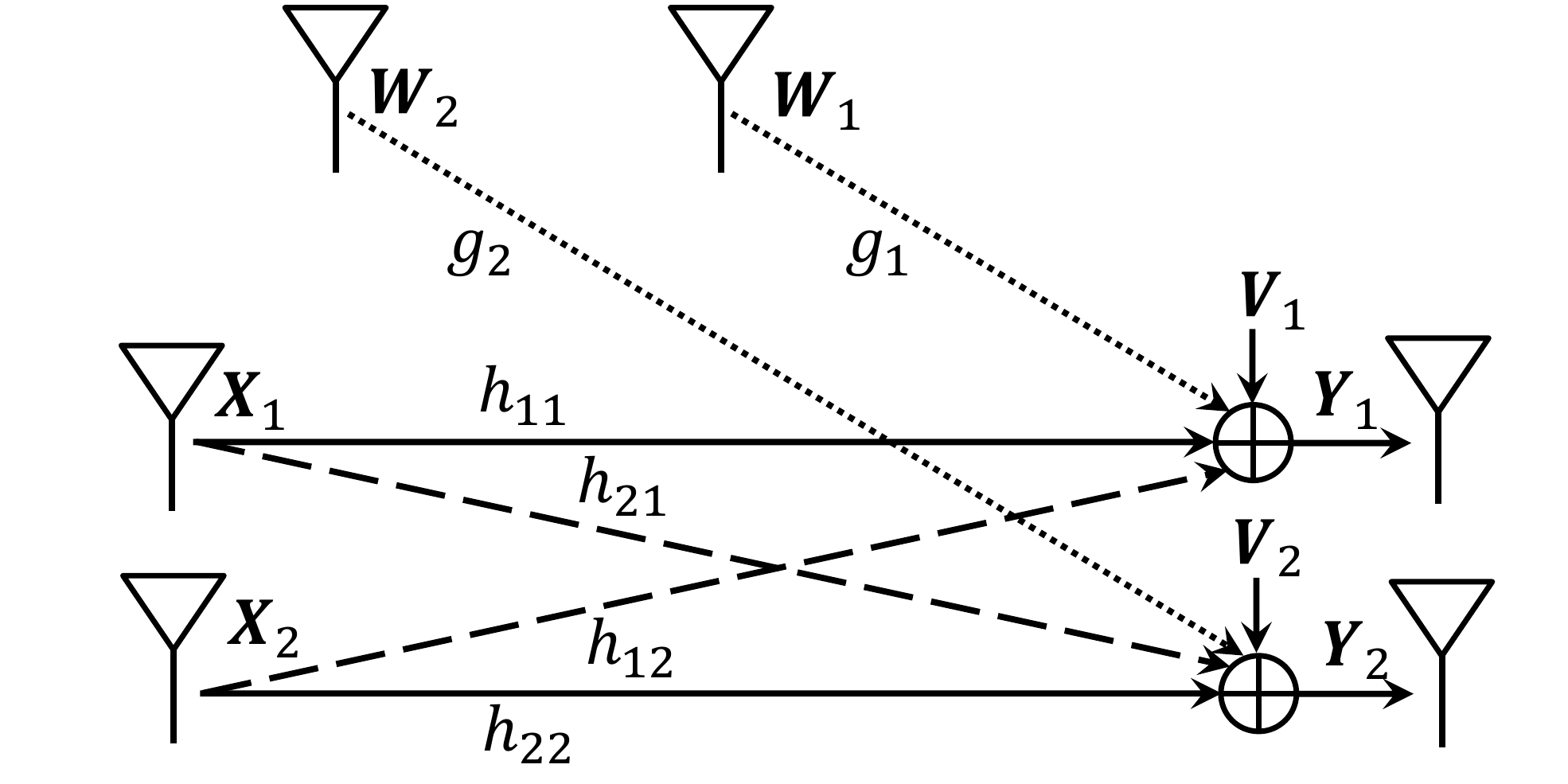}
    \vspace{-1em}\caption{Two-user Gaussian Interference Channel with Two Independent Jammers.}
    \label{fig:Fig1Label}
\end{figure}

The Gaussian interference channel with two independent jammers is shown in Fig.~\ref{fig:Fig1Label}, in which two users send their messages to their own receivers in the presence of one or two jammers. The jammers are assumed not to have any information about the user's signals (but know the code). In particular, the received signals are given by
\begin{align}\label{eq:1}
\begin{split}
\bY_1 = h_{11}\bX_1 + h_{12}\bX_2 + g_1 \bW_1 + \bV_1\\
\bY_2 = h_{21}\bX_1 + h_{22}\bX_2 + g_2 \bW_2 + \bV_2
\end{split}
\end{align} 
where $\bX_1$ and $\bX_2$ are $n$-length vectors representing the user's signals, $\bW_1$ and $\bW_2$ are the independent adversarial jammer signals, $h_{ij}$ and $g_i$ for $i,j\in\{1,2\}$ are the channel gains, and $\bV_i$ is the $n$-length noise vector distributed as a sequence of i.i.d. zero mean Gaussian random variables with variance $\sigma^2$ which is independent of $\bX_1$, $\bX_2$, $\bW_1$ and $\bW_2$.

The transmitter and jammer signals are constrained to satisfy power constraints $\|\bX_i\|^2\le~nP_i$ and $\|\bW_i\|^2 \!\le\! n\Lambda$, for $i \!= \!1,2$, respectively. We define the received signal-to-noise and interference-to-noise ratios as $S_1 = h_{11}^2P_1/\sigma^2$, $S_2 = h_{22}^2P_2/\sigma^2$, $I_1 = h_{12}^2P_2/\sigma^2$ and $I_2 = h_{21}^2P_1/\sigma^2$. We also denote the jammer-to-noise ratios as $J_1=g_1^2\Lambda/\sigma^2$ and $J_2=g_2^2\Lambda/\sigma^2$. We assume that the transmitters and receivers know the signal-to-noise and interference-to-noise ratios, but they need not know the jammer-to-noise ratios. However, we require small probability of error only when the jammer-to-noise ratios do not exceed $J_1$,$J_2$; thus the code is independent of the jammer's power up to a point, and beyond that it may fail to decode correctly.

A $\left(2^{nR_1},2^{nR_2},n\right)$ deterministic code is given by: 
\begin{itemize}
  \item Message sets $\mathcal{M}_1=[2^{nR_1}]$ and $\mathcal{M}_2=[2^{nR_2}]$,
  \item Encoding functions $\bx_i:\mathcal{M}_i\to \mathbb{R}^n$ for $i=1,2$, and 
	\item Decoding functions $\phi_i:\mathbb{R}^n\to \mathcal{M}_i$ for $i=1,2$.
\end{itemize}

For $i \!=\! 1,2$, the message $M_i$ is chosen uniformly from the set $\mathcal{M}_i$, and each transmitter encodes its own message to $\bX_i$. At each receiver, the received signal $\bY_i$ is decoded by function $\phi_i$ to $\hat{M}_i \!=\! \phi_i(\bY_i)$. The average probability of error $P_e^{(n)}$ is now given by the probability that $(\hat{M}_1,\hat{M}_2)\!\ne\!(M_1,M_2)$, maximized over all possible choices of jammers' sequences $\bW_1$ and $\bW_2$. A rate pair $(R_1,R_2)$ is \emph{achievable} if there exists a sequence of $\left(2^{nR_1},2^{nR_2},n\right)$ codes where $\underset{n \rightarrow \infty}{\lim}P_e^{(n)} \!=\! 0$. The capacity region $\mathscr{C}$ is the closure of the set of all achievable rate pairs $(R_1,R_2)$. 

\begin{figure}[t]
\centering
    \includegraphics[width=0.5\linewidth]{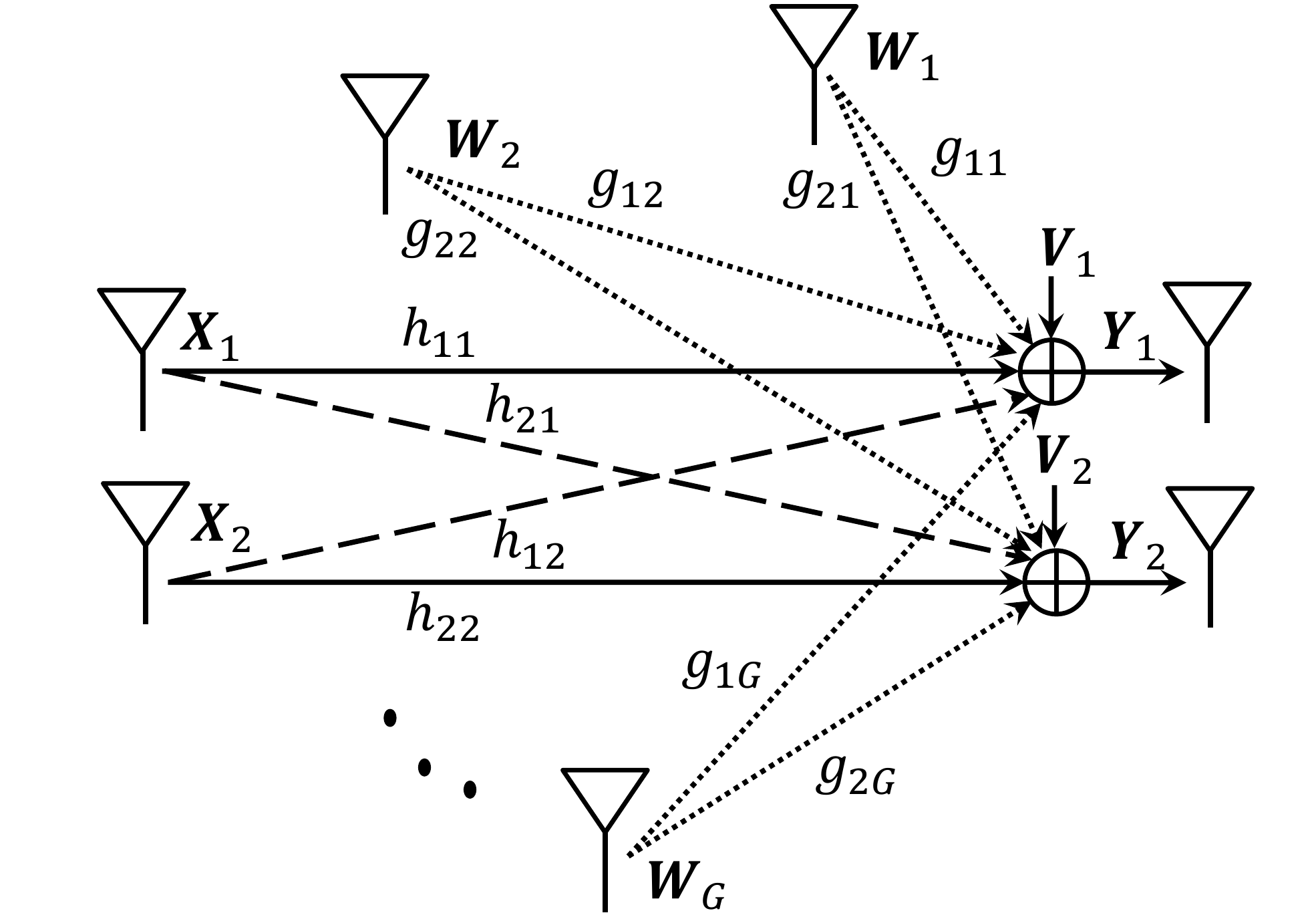}
    \vspace{-1em}\caption{Two-user Gaussian Interference Channel with Two Independent Cross Jammers.}
    \label{fig:Fig22Label}
\end{figure}
\subsection{Generalized Jamming Model}
Generally speaking, if there are $G$ jammers ($G\geq 1$) with the cross matrix $\mathscr{G}$
\begin{align}
	\mathscr{G} = \begin{bmatrix}
    g_{11} & g_{12} & g_{13} & \dots & g_{1G} \\
    g_{21} & g_{22} & g_{23} & \dots & g_{2G}
    \end{bmatrix}
\end{align}
as shown in Fig.~\ref{fig:Fig22Label}, then the received signals at each decoder are given by
\begin{align}
\begin{split}
\bY_1 = h_{11}\bX_1 + h_{12}\bX_2 + g_{11} \bW_1 +g_{12} \bW_2 +\ldots +g_{1G} \bW_G + \bV_1\\
\bY_2 = h_{21}\bX_1 + h_{22}\bX_2 + g_{21} \bW_1 +g_{22} \bW_2 +\ldots +g_{2G} \bW_G + \bV_2
\end{split}
\end{align}
where $\|\bW_i\|^2 \le n\Lambda$ for $i = 1,2,\ldots,G$. This includes the case where there is only one jammer ($G=1$). We refer the capacity region of this channel as $\mathscr{C}_{G}$, and state the following proposition for the relation between $\mathscr{C}_{G}$ and $\mathscr{C}$. Indeed, the capacity region depends only on the received signal at each decoder and not the number of jammers.
\begin{proposition}\label{Prp2}
We have $\mathscr{C}_{G}=\mathscr{C}$ as long as 
\begin{align}\label{power-condition}
\begin{split}
|g_{11}|+ |g_{12}|+ \ldots+ |g_{1G}|=|g_1|\\
|g_{21}|+ |g_{22}|+ \ldots+ |g_{2G}|=|g_2|
\end{split}
\end{align}
where the jammer-to-noise ratios are then given by $J_{1} = \left(g_{11}+ g_{12}+ \ldots+ g_{1G}\right)^2\Lambda/\sigma^2$ and $J_{2} = \left(g_{21}+ g_{22}+ \ldots+ g_{2G}\right)^2\Lambda/\sigma^2$.
\end{proposition}
The proof is provided in Appendix \ref{Apn0}.

\section{Main Results}\label{3}

In this section, we present inner and outer bounds on the capacity region $\mathscr{C}$ (two-user GIC with two independent jammers). Before stating the main results, we define regions $\mathscr{R}_o(S_1,S_2,I_1,I_2)$ and $\mathscr{R}_i(S_1,S_2,I_1,I_2)$ as the previously-derived outer and inner bounds respectively for the GIC with no jammer; namely $\mathscr{R}_o$ is the outer bound of \cite{Etkin2008}, and $\mathscr{R}_i$ is the Han-Kobayashi inner bound \cite{Han1981}. When we write an expression with $i$ and $j$, we mean for it to hold for both $(i,j)=(1,2)$ and $(i,j)=(2,1)$.

Define $\mathscr{R}_o(S_1,S_2,I_1,I_2)$ as the set of rate pairs $(R_1,R_2)$ such that
\begin{align*}
\allowdisplaybreaks
R_i &\le C\left(S_i\right)\\
R_i + R_j &\le C\left({\textstyle\frac{S_i}{1+I_j}}\right) + C\left(I_j+S_j \right)\\
R_1 + R_2 &\le C\left({\textstyle\frac{S_1 + I_1 + I_1 I_2}{1+I_2}}\right) + C\left({\textstyle\frac{S_2 + I_2 + I_1 I_2}{1+I_1}}\right)\\
2R_i + R_j &\le C\left({\textstyle\frac{S_i}{1+I_j}}\right) + C\left(S_i+I_i\right)\! + C\left({\textstyle\frac{S_j + I_j + I_i I_j}{1+I_i}}\right).
\end{align*}
Define $\mathscr{R}_i(S_1,S_2,I_1,I_2)$ as the set of rate pairs $(R_1,R_2)$ such that
\begin{align*}
\begin{split}
R_i &< C\left({\textstyle\frac{S_i}{1+\alpha_jI_i}}\right) \\
R_i+R_j &< C\left(\textstyle{\frac{S_i+\bar{\alpha}_j I_i}{1+\alpha_j I_i}}\right) + C\left(\textstyle{\frac{\alpha_j S_j}{1+\alpha_i I_j}}\right) \\
R_1+R_2 &< C\left(\textstyle{\frac{\alpha_1 S_1+\bar{\alpha}_2 I_1}{1+\alpha_2 I_1}}\right) + C\left(\textstyle{\frac{\alpha_2 S_2+\bar{\alpha}_1 I_2}{1+\alpha_1 I_2}}\right)\\
2R_i+R_j &< C\left(\textstyle{\frac{S_i+\bar{\alpha}_j I_i}{1+\alpha_j I_i}}\right) \!+ \!C\left(\textstyle{\frac{\alpha_i S_i}{1+\alpha_j I_i}}\right) \!+  \! C\left(\textstyle{\frac{\alpha_j S_j+\bar{\alpha}_i I_j}{1+\alpha_i I_j}}\right)
\end{split}
\end{align*}
for some $\alpha_i$ in $[0,1]$ where $\alpha_i$ implies the portion of the private message power in the Han-Kobayashi inner bound proof at user $i$. Note that in the Han-Kobayashi inner bound proof encoder $i$ divides the message $m_i$ into private message $m_{ip}$ and common message $m_{ic}$ with power $\alpha_i P_i$ and $\bar{\alpha_i} P_i$ respectively.

Define $S_i^\prime = \frac{S_i}{1+J_i}$ and $I_i^\prime = \frac{I_i}{1+J_i}$. We now state our main outer and inner bounds.

\begin{theorem}[Outer Bound]\label{Thrm1}$\mathscr{C}\! \!\subseteq \!\mathscr{R}_o(S_1',S_2',I_1',I_2')$. Moreover, if $S_1\!\!\le\! J_1$ or $S_2\!\!\le\! J_2$, then $\mathscr{C}=\emptyset$.
\end{theorem}

\begin{theorem}[Inner Bound]\label{Thrm2} Assume $S_i>J_i$ for $i=1,2$. Let $\tilde{\mathscr{R}}_i(S_1',S_2',I_1',I_2')$ be the subset of rate pairs in $\mathscr{R}_i(S_1',S_2',I_1',I_2')$ achieved by $\alpha_i\in[0,1]$ satisfying 
\begin{align}\label{alpha_condition}
	\alpha_i S_i+\bar\alpha_j I_i>J_i \text{ for } (i,j)=(1,2),(2,1).
\end{align}
Then $\tilde{\mathscr{R}}_i(S_1',S_2',I_1',I_2')\subseteq\mathscr{C}$.
\end{theorem}

\begin{figure}[p]
\centering
\subfloat[$S=4$, $I=3$, and $J$ between 0 and 5. For these parameters the bound $\mathscr{R}_i(S',S',I',I')$ is identical to our inner bound if the jammer-to-noise ratio is less than 3.2.]{
	\label{subfiga}
	\includegraphics[width=0.7\textwidth]{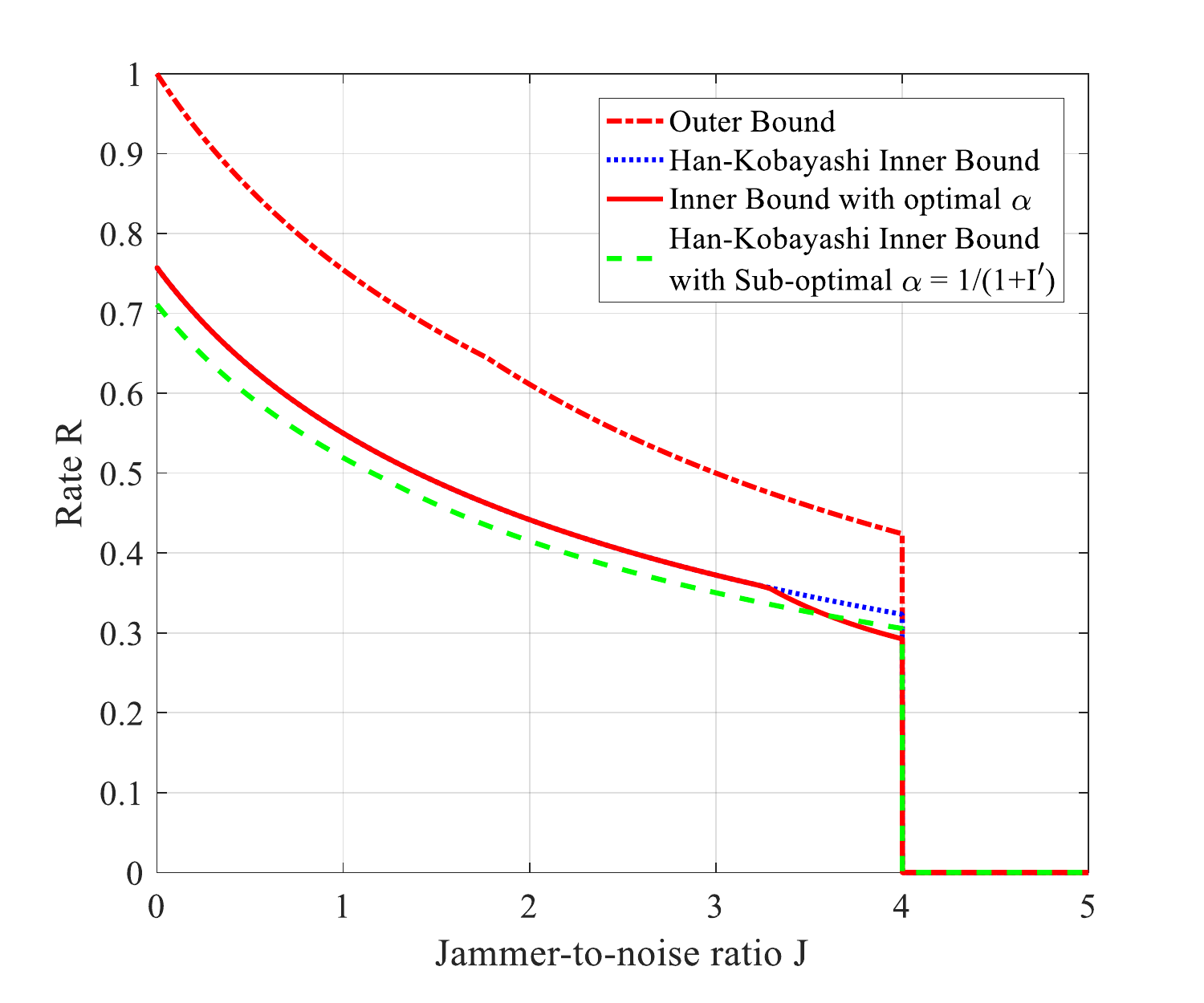} } 
 
\subfloat[ $S=4$, $J=3.5$, and $I$ between 0 and 10. The bound $\mathscr{R}_i(S',S',I',I')$ is identical to our inner bound for weak and strong interference.]{
	\label{subfigb}
	\includegraphics[width=0.45\textwidth]{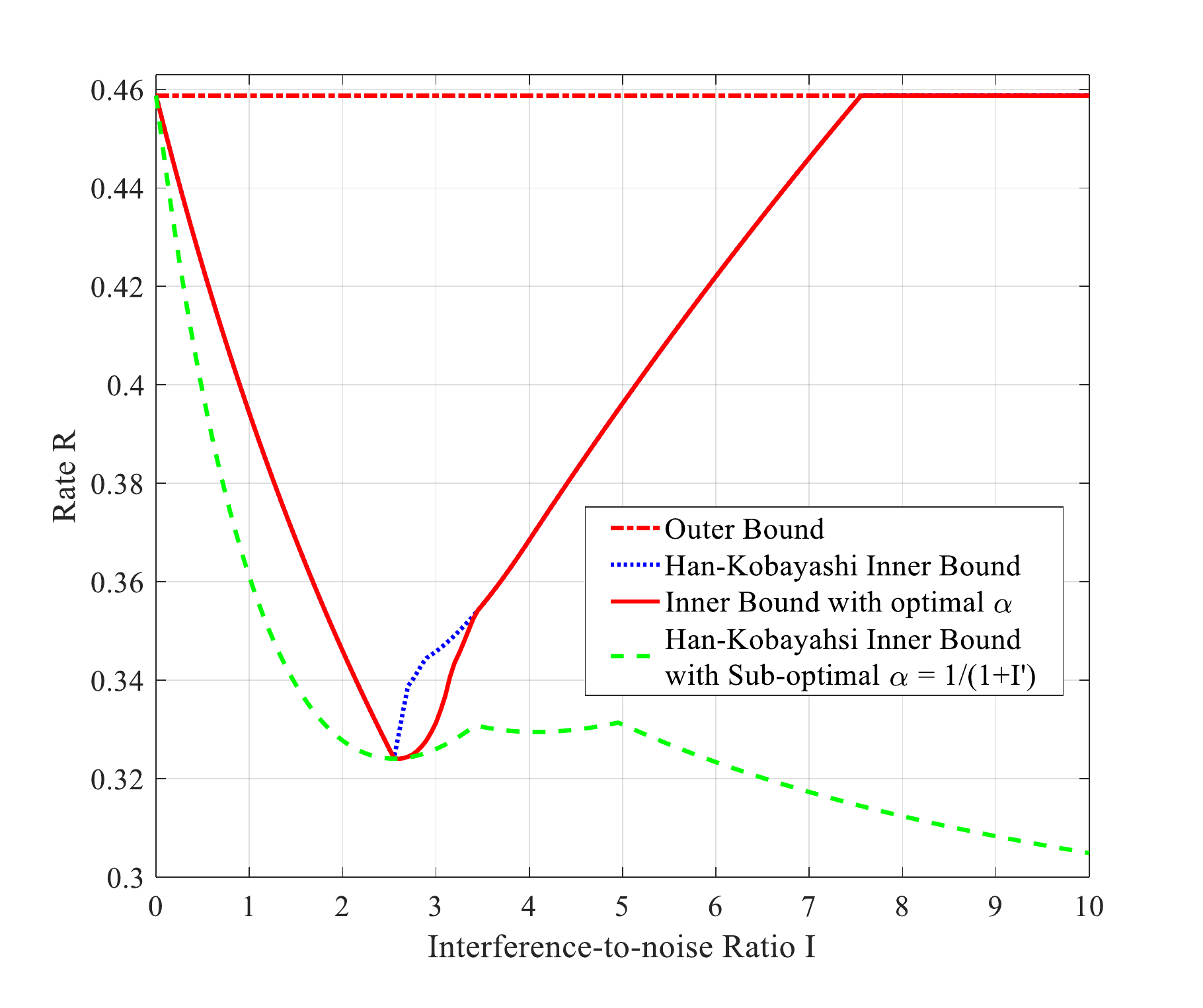} } 
	\hspace{2em}
\subfloat[$S=10$, $J=3.5$, and $I$ between 0 and 40. The bound $\mathscr{R}_i(S',S',I',I')$ is identical to our inner bound for high signal-to-noise ratio $S=10$.]{
	\label{subfigc}
	\includegraphics[width=0.45\textwidth]{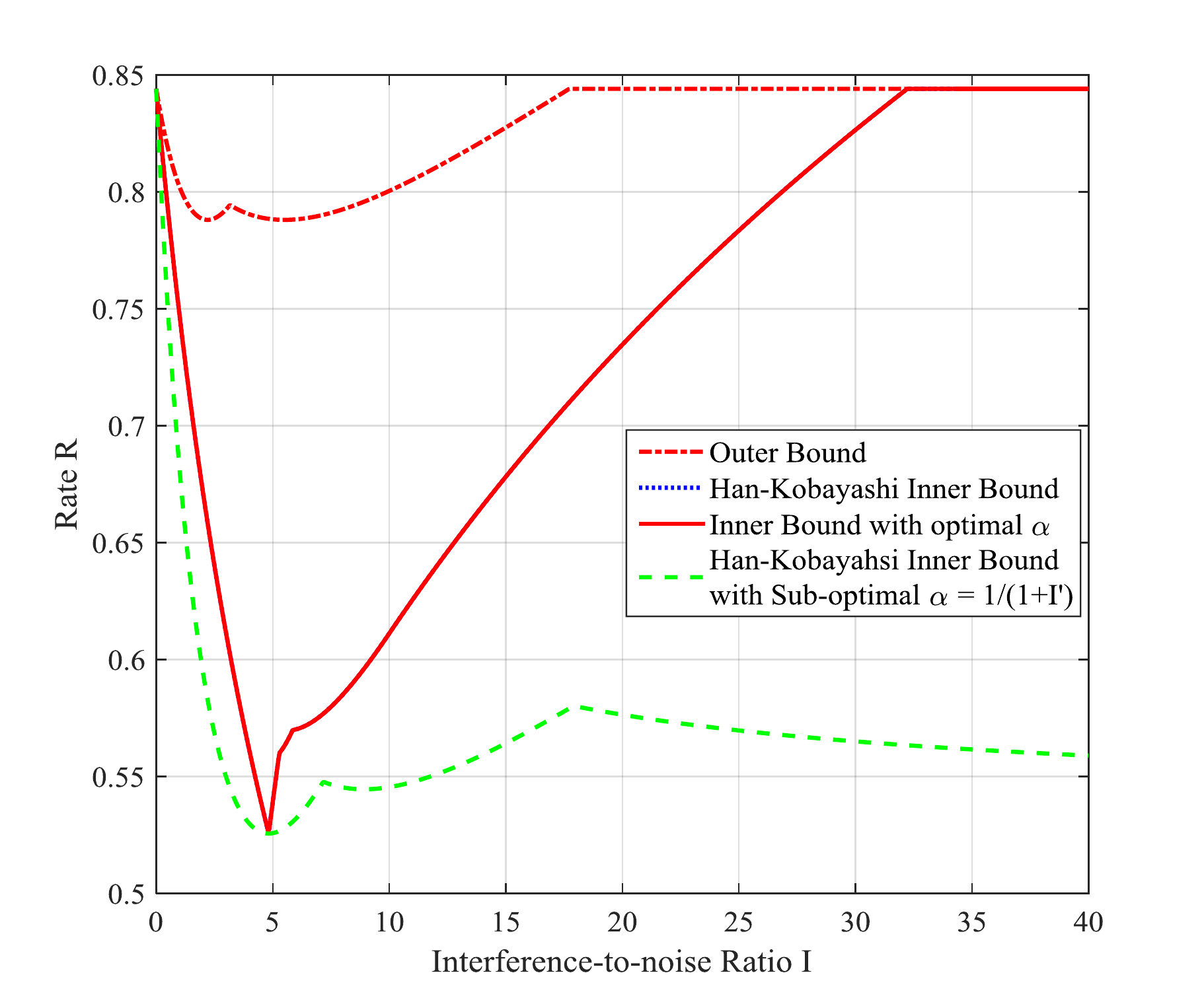}} 
\caption{Bounds on the symmetric capacity $\mathscr{C}_{\text{sym}}(S,I,J)$ for $S_1=S_2=S$, $I_1=I_2=I$, and $J_1=J_2=J$. In addition to our inner $\tilde{\mathscr{R}}_i(S',S',I',I')$ and outer bounds $\mathscr{R}_o(S',S',I',I')$, also shown the bound $\mathscr{R}_i(S',S',I',I')$ and shown $\mathscr{R}_i(S',S',I',I')$ with sub-optimal $\alpha =\frac{1}{1+I'}$.}
\label{fig:Fig2Label}
\end{figure}

\section{Discussion and Numerical Results}\label{4}

Note that the inner bound differs from $\mathscr{R}_i(S_1',S_2',I_1',I_2')$ only when the optimal $\alpha_i$ parameters do not satisfy \eqref{alpha_condition}. However, in several regimes of interest, this constraint is not active. For example, if the channel has weak interference, in the sense of \cite[eq (6.8)]{ElGamal} $\sqrt{\frac{I_j'}{S_i'}} (1 + I_i')\leq \rho_i(1 - \rho_j)$ for some $\rho_1,\rho_2\in[0,1]$ and $(i,j)=(1,2),(2,1)$ then treating interference as noise is optimal for the sum-rate \cite[Theorem 6.3]{ElGamal}. Treating interference as noise corresponds to $\alpha_1=\alpha_2=1$, under which \eqref{alpha_condition} holds. Therefore, in the weak interference regime, our inner bound matches $\mathscr{R}_i(S_1',S_2',I_1',I_2')$, and it also achieves the exact sum-rate capacity. On the other hand, when the channel has strong interference in both users $I_2'\geq S_1'$ and $I_1'\geq S_2'$, by choosing $\alpha_1=\alpha_2=0$ each transmitter only sends its own common message, and both messages can be decoded at each receiver. Therefore, \eqref{alpha_condition} holds if we have $I_1>J_1$ and $I_2>J_2$. Thus, we obtain the exact capacity region for the strong interference regime \cite[Theorem 6.2]{ElGamal}. 

In \cite[Theorem 6.6]{ElGamal}, it is shown that using Han-Kobayashi inner bound with sub-optimal choices $\alpha_1=\frac{1}{1+I'_2}$ and $\alpha_2=\frac{1}{1+I'_1}$ yields an inner bound that is always within half a bit of the outer bound. Therefore, if $\alpha_1=\frac{1}{1+I'_2}$ and $\alpha_2=\frac{1}{1+I'_1}$ satisfy our conditions in \eqref{alpha_condition} then our inner bound is guaranteed to be within half a bit of our outer bound; that is, if $J_1<\frac{S_1}{1+I'_2}+\frac{I_1^2}{1+I'_1}$ and $J_2<\frac{S_2}{1+I'_1}+\frac{I_2^2}{1+I'_2}$, our inner and outer bounds are within half a bit. 

Now, consider the symmetric case; i.e. $S_1=S_2=S$, $I_1=I_2=I$, $J_1=J_2=J$, and $R_1=R_2=R$. Clearly in this case it is optimal to choose $\alpha_1=\alpha_2=\alpha$ for inner bound. Define the symmetric capacity of the channel as $C_{\text{sym}}(S,I,J) =\max\{R: (R, R) \in \mathscr{C}\}$. We illustrate the bounds for $C_{\text{sym}}(S,I,J)$ in Fig.~\ref{fig:Fig2Label} including our outer bound $\mathscr{R}_o(S',S',I',I')$, our inner bound $\tilde{\mathscr{R}}_i(S',S',I',I')$ with optimal $\alpha$, the Han-Kobayashi inner bound with the noise variance increased by the received power of the jammer $\mathscr{R}_i(S',S',I',I')$ and the latter with sub-optimal $\alpha=\frac{1}{1+I'}$.
Note that we are in the strong interference regime only if $I'\ge S'$.

\begin{figure}[t]
\centering
\subfloat[DoF for $\delta=\frac{1}{4}$ and $\beta$ between $0,2$.]{
	\label{DoFsubfiga}
	\includegraphics[width=0.45\textwidth]{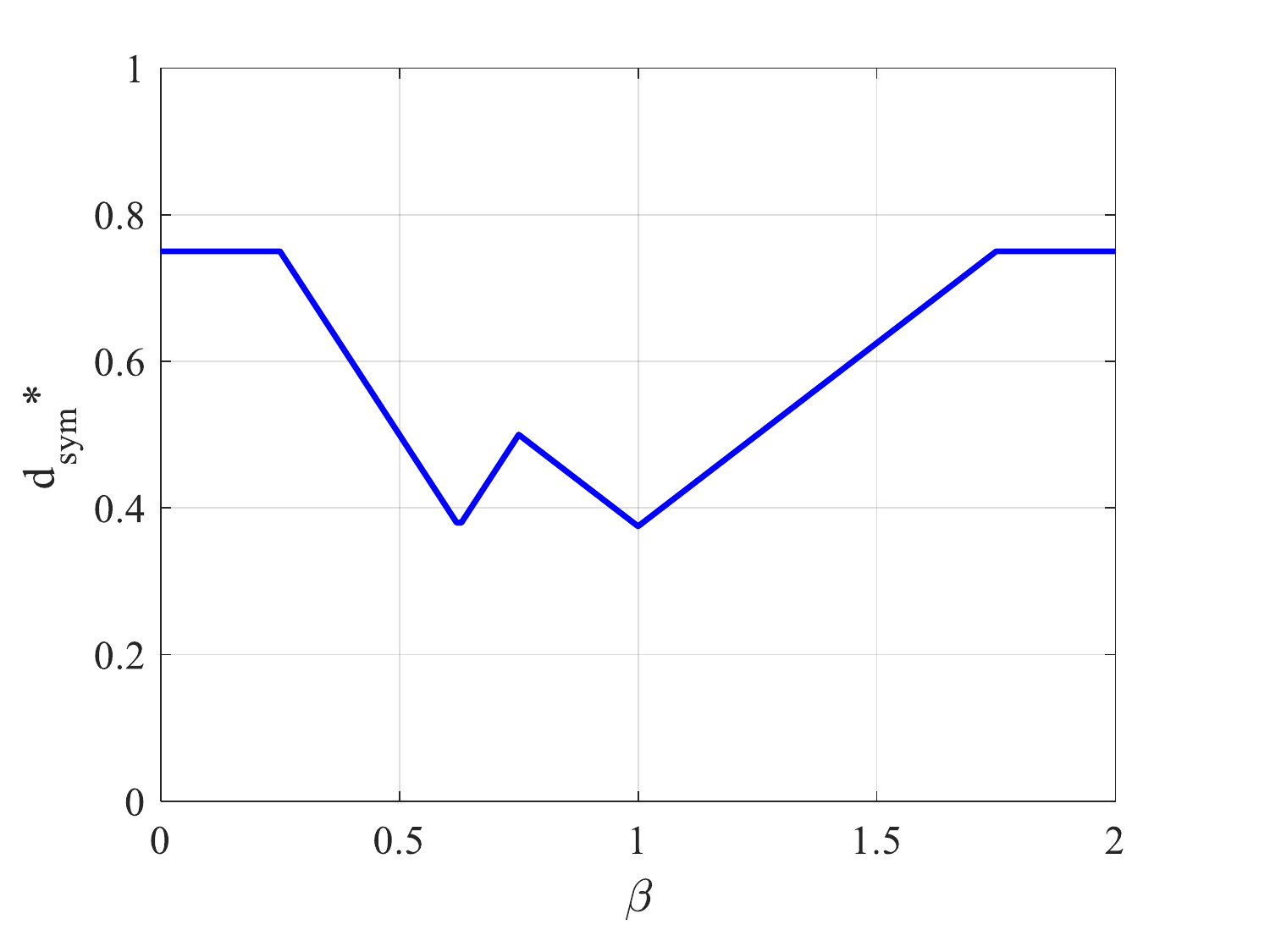} } 
	\hspace{1.5em}
\subfloat[DoF for $\beta=0.7$ and $\delta$ between $0,2$.]{
	\label{DoFsubfigb}
	\includegraphics[width=0.45\textwidth]{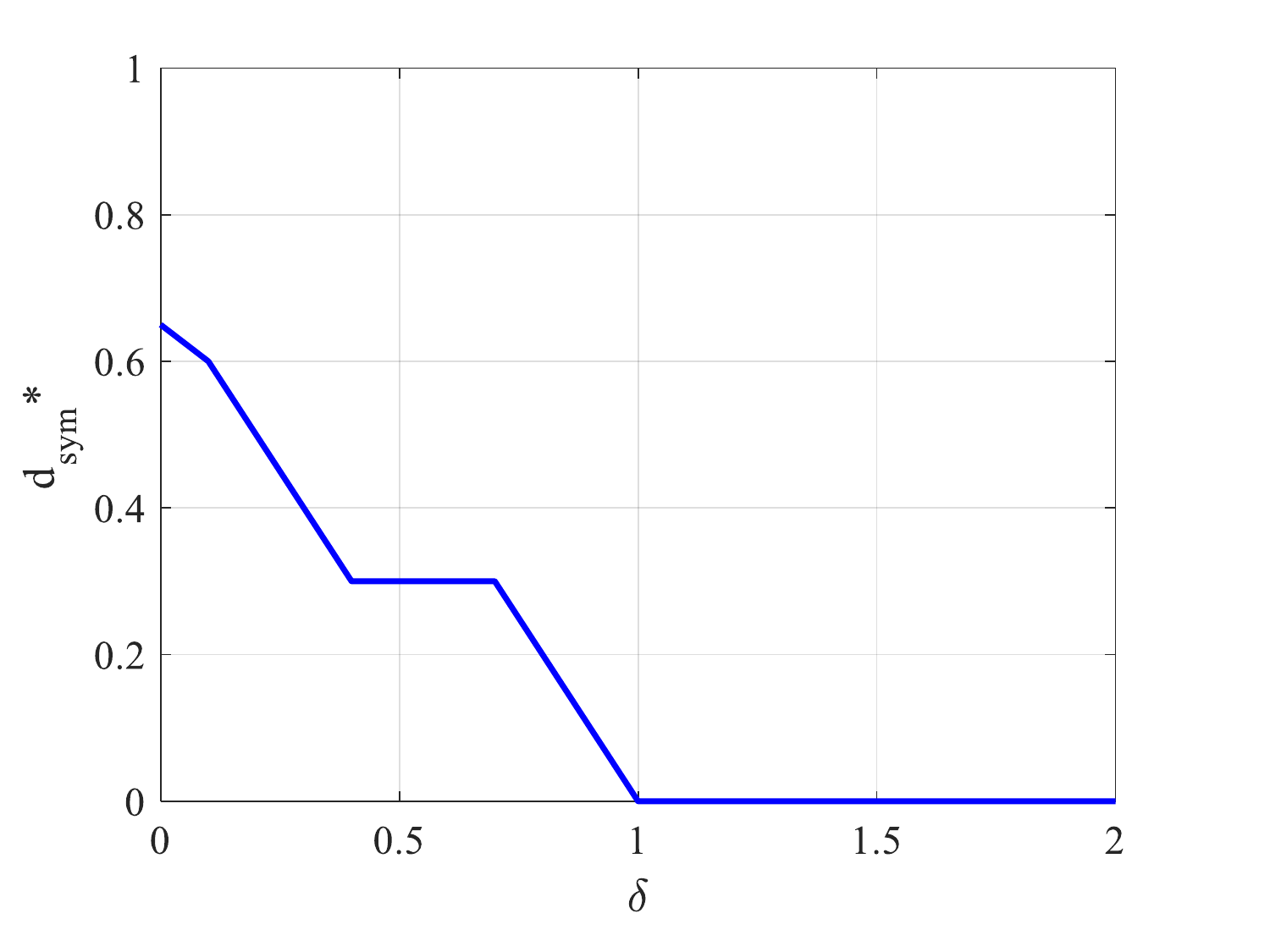}} 
\caption{Symmetric degrees of freedom for the GIC.}
\label{fig:DoF}
\end{figure}

Define the normalized symmetric capacity as $d_{\text{sym}} = \frac{C_{\text{sym}}(S,I,J)}{C(S)}$. Then the symmetric degrees of freedom (DoF) $d^*_{\text{sym}}$ is given by
\begin{align}
	d^*_{\text{sym}}(\beta, \delta) = \lim_{S\to \infty} \frac{C_{\text{sym}}(S,S^\beta,S^\delta)}{C(S)}.
\end{align}
By substituting $I=S^\beta$ and $J=S^\delta$ in our outer and inner bounds $\mathscr{R}_{o}(S',S',I',I')$ and $\tilde{\mathscr{R}}_{i}(S',S',I',I')$, we find the upper bound for $C_{\text{sym}}(S,S^\beta,S^\delta)$ given by
\begin{multline}
C_{\text{sym}}(S,S^\beta,S^\delta)\leq \max \{R: (R,R)\in \mathscr{R}_{o}(S',S',I',I')\}=\\ \min\bigg\{ C\left(\frac{S}{1+S^\delta}\right), \frac{1}{2}C\left({\frac{S}{1+S^\delta+S^\beta}}\right) + \frac{1}{2}C\left(\frac{S+S^\beta}{1+S^\delta} \right), C\left({\textstyle\frac{S + S^\beta + \frac{S^{2\beta}}{1+S^\delta}}{1+S^\delta+S^\beta}}\right) ,\\
 \frac{1}{3} C\left({\frac{S}{1+S^\delta+S^\beta}}\right) + \frac{1}{3} C\left({\frac{S+S^\beta}{1+S^\delta}}\right)+ \frac{1}{3}C\left({\textstyle\frac{S + S^\beta + \frac{S^{2\beta}}{1+S^\delta}}{1+S^\delta+S^\beta}}\right)\bigg\},\label{upperB}
\end{multline}
and the lower bound for $C_{\text{sym}}(S,S^\beta,S^\delta)$ given by
\begin{multline}
C_{\text{sym}}(S,S^\beta,S^\delta)\geq \max \{R: (R,R)\in \tilde{\mathscr{R}}_{i}(S',S',I',I')\} =\\ \max_{\alpha: \alpha S+\bar{\alpha}S^\beta>S^\delta} \bigg\{\min\bigg\{C\left({\frac{S}{1+S^\delta+\alpha S^\beta}}\right),
 \frac{1}{2}C\left(\frac{S+\bar{\alpha} S^\beta}{1+S^\delta+\alpha S^\beta}\right) + \frac{1}{2}C\left(\frac{\alpha S}{1+S^\delta+\alpha S^\beta}\right),\\
C\left(\frac{\alpha S+\bar{\alpha} S^\beta}{1+S^\delta+\alpha S^\beta}\right),
\frac{1}{3}C\left(\frac{S+\bar{\alpha} S^\beta}{1+S^\delta+\alpha S^\beta}\right) + \frac{1}{3}C\left(\frac{\alpha S}{1+S^\delta+\alpha S^\beta}\right) + \frac{1}{3}  C\left(\frac{\alpha S+\bar{\alpha} S^\beta}{1+S^\delta+\alpha S^\beta}\right) \bigg\}\bigg\}.\label{lowerB}
\end{multline}
We may further lower bound the symmetric capacity by choosing $\alpha=\frac{1}{1+I'}=\frac{1+S^\delta}{1+S^\delta+S^\beta}$ as long as this choice satisfies \eqref{alpha_condition}. In particular, we claim that this value of $\alpha$ always satisfies \eqref{alpha_condition} for sufficiently large $S$. We show this by substituting this value of $\alpha$ to find
\begin{align}
\alpha S+\bar{\alpha} S^\beta
	&=\frac{1+S^\delta}{1+S^\delta+S^\beta} S+\frac{S^\beta}{1+S^\delta+S^\beta} S^\beta \\
	&=\frac{S+S^{1+\delta}+S^{2\beta}}{1+S^\delta+S^\beta}.\label{sym_condition1}
\end{align}
Since the capacity region is empty when $S\le J=S^\delta$, it suffices to consider \eqref{sym_condition1} only for $\delta<1$. The dominant power of $S$ in \eqref{sym_condition1} is given by
\begin{equation}
\max\{1+\delta,2\beta\}-\max\{\delta,\beta\}
>\max\{2\delta,2\beta\}-\max\{\delta,\beta\}
=\max\{\delta,\beta\}
\ge \delta.
\end{equation}
Therefore, \eqref{sym_condition1} is larger than $J=S^\delta$ for sufficiently large $S$, thus $\alpha=\frac{1}{1+I'}$ satisfies \eqref{alpha_condition}.
Now, we may substitute this choice of $\alpha$ into \eqref{lowerB}, and take the limits of \eqref{lowerB} and \eqref{upperB} as $S\to \infty$. Therefore, we find that the symmetric DoF is given by 
\begin{align}
	d^*_{\text{sym}}(\beta,\delta) = \min\left\{\max\{0,1-\delta\},\max\left\{0, 1-\beta, \beta-\delta\right\}, \max\left\{0, 1-\frac{\beta}{2}-\frac{\delta}{2},\frac{\beta}{2}-\frac{\delta}{2} \right\}  \right\}
\end{align}
which is illustrated for fixed $\delta=1/4$ in Fig.~\ref{DoFsubfiga} and fixed $\beta=0.7$ in Fig.~\ref{DoFsubfigb}. Note that for the interference channel with no jammer \cite[p. 153]{ElGamal}, the DoF exhibits a ``W'' shape for a fixed $\delta=\frac{\log J}{\log S}$.

\section{Proof of Outer Bound}\label{5}

Consider a sequence of $(2^{nR_1},2^{nR_2},n)$ codes with vanishing probability of error. Since these codes must function for arbitrary jamming signals, we may derive an outer bound by assuming the jammers transmit Gaussian noise with variance $\Lambda$. Thus, we follow the outer bound for the GIC with no jammer \cite[Chapter 6.7.2, p. 151]{ElGamal} and the noise power $\sigma^2+g_i^2\Lambda$. This yields the outer bound $\mathscr{R}_o(S_1',S_2',I_1',I_2')$.

Moreover, if $J_1\ge S_1$, based on the assumption that the jammer knows the code, the jammer can choose an arbitrary message $\tilde{m}_1$ and transmit a scaled form of the corresponding codeword $\bw_1 = \bx_1({\tilde{m}_1})h_{11}/g_1$. Given $\bY_1=h_{11}\bx_1(m_1)+h_{12}\bx_2(m_2)+h_{11}\bx_1({\tilde{m}_1})+\bV_1$, decoder 1 cannot decode the message since it does not know whether the true message is $m_1$ or $\tilde{m}_1$. The same scenario can happen for decoder 2 if $J_2\ge S_2$. This attack constitutes AVC symmetrization.

\section{Proof of Inner Bound}\label{6}

Our inner bound proof is a generalization of the Han-Kobayashi bound \cite[Chapter 6.5.1, p. 144]{ElGamal}. Using rate splitting, we represent message $m_i$ from user $i$ for $i=1,2$, by independent common message $m_{ic}$ at rate $R_{ic}$ and private message $m_{ip}$ at rate $R_{ip}$ such that $R_i = R_{ic}+R_{ip}$. Thus, each receiver will decode its own common and private messages and the common message of the interfering user. Assuming $S_i>J_i$ for $i=1,2$, we show that $(R_{1c},R_{1p},R_{2c},R_{2p})$ is achievable if
\begin{align}\label{eq:2}
\begin{split}
R_{ip} &< C\left(\textstyle{\frac{\alpha_i S_i}{1+J_i+\alpha_j I_i}}\right)\\
R_{ip} + R_{ic} &< C\left(\textstyle{\frac{S_i}{1+J_i+\alpha_j I_i}}\right)\\
R_{ip} + R_{jc} &< C\left(\textstyle{\frac{\alpha_i S_i+\bar{\alpha}_j I_i}{1+J_i+\alpha_j I_i}}\right)\\
R_{ip} + R_{ic} + R_{jc} &< C\left(\textstyle{\frac{S_i + \bar{\alpha}_j I_i}{1+J_i+\alpha_j I_i}}\right)
\end{split}
\end{align}
for some $\alpha_i\in[0,1]$ satisfying $ \alpha_i S_i+\bar\alpha_j I_i>J_i$, and again the above holds for $(i,j)=(1,2)$ and $(i,j)=(2,1)$. This achieves the region $\tilde{\mathscr{R}}_i$ by substituting $R_1=R_{1c}+R_{1p}$ and $R_2=R_{2c}+R_{2p}$, and applying the Fourier-Motzkin procedure to eliminate $R_{ic}$ and $R_{ip}$.

Before proceeding to the proof, we first define the following typical set for Gaussian random variables $X_1,\ldots,X_k$ as:
\begin{multline}
\T(X_1,\ldots,X_k)\\
=\left\{(\bx_1,\ldots,\bx_k):\bbE(X_iX_j)-\eps\le \frac{1}{n}\langle \bx_i,\bx_j\rangle \le \bbE(X_iX_j)+\eps \text{ for all }i,j\in[1:k]\right\}.\label{TypicalSet}
\end{multline}

\emph{Codebook generation:} Fix $\alpha_1, \alpha_2 \in [0,1]$ and $\gamma>0$. For $i=1,2$, we generate $2^{nR_{ic}}$ i.i.d zero mean Gaussian sequences $\bX_{ic}(m_{ic})$ with variance $(1-\gamma)\bar{\alpha}_iP_i$ for each $m_{ic} \in [2^{nR_{ic}}]$. Also, for each $m_{ic}\in[2^{nR_{ic}}]$, generate $2^{nR_{ip}}$ i.i.d. zero mean Gaussian sequences $\bX_{ip}(m_{ic}, m_{ip})$ with variance $(1-\gamma)\alpha_i P_i$ for each $m_{ip} \in [2^{nR_{ip}}]$ for $i=1,2$.

\emph{Encoding:} For $i=1,2$, write message $m_i$ as $(m_{ic},m_{ip})$ where $m_{ic}\in[2^{nR_{ic}}]$ and $m_{ip}\in[2^{nR_{ip}}]$. Transmitter $i$ sends $\bX_i=\bX_{ic}(m_{ic})+\bX_{ip}(m_{ic},m_{ip})$ if its power is less than $P_i$, otherwise it sends zero.  

\emph{Decoding:} We describe the decoding procedure for receiver 1; that of receiver 2 is similar. 
First, let 
\be
\mathscr{S} =\left\{(m_{1c},m_{1p},m_{2c}):
(\bx_{1c}(m_{1c}), \bx_{1p}(m_{1c},m_{1p}),\bx_{2c}(m_{2c}),\by_1)\in \bigcup \, \T(X_{1c},X_{1p},X_{2c},Y_1) \right\}\label{setS}
\ee
where the union is over all joint Gaussian distributions $X_{1c},X_{1p},X_{2c},Y_1$ such that $(X_{1c},X_{1p},X_{2c},\allowbreak Y_1-h_{11} X_{1c}-h_{11} X_{1p} - h_{12} X_{2c})$ are mutually independent.

Given $\by_1$, decoder 1 finds
\begin{align}
	(\hat{m}_{1c},\hat{m}_{1p},\hat{m}_{2c})= \argmin_{(m_{1c},m_{1p},m_{2c}) \in  \mathscr{S}}
 \left\|\by_1-h_{11}\bx_{1c}( m_{1c})-h_{11}\bx_{1p}(m_{1c}, m_{1p})-h_{12}\bx_{2c}(m_{2c})\right\|.
\end{align}
If there is more than one minimum, choose between them arbitrarily. The decoder then outputs the message estimate $\hat{m}_1=(\hat{m}_{1c},\hat{m}_{1p})$.

\emph{Analysis of the probability of error:} Assume the two users send messages $\big(( M_{1c},$ $M_{1p}),( M_{2c}, \allowbreak M_{2p})\big)$. We will obtain the average probability of error for decoder 1 and similarly generalize the results for decoder 2. Define the error event 
\begin{align}
	\mathcal{E}_0=\left\{(M_{1c},M_{1p},M_{2c})\notin \mathscr{S}\right\}.
\end{align}
To consider error events in which a false message set appears correct, we define the set
\begin{multline}
\mathscr{T}=\big\{(m_{1c},m_{1p},m_{2c})\in\mathscr{S}:
\|\bY_1-h_{11}\bx_{1c}( m_{1c})-h_{11}\bx_{1p}(m_{1c}, m_{1p})-h_{12}\bx_{2c}( m_{2c})\|^2 \\
\leq \|\bY_1-h_{11}\bx_{1c}( M_{1c})-h_{11}\bx_{1p}(M_{1c}, M_{1p})-h_{12}\bx_{2c}( M_{2c})\|^2 \big\}.
\end{multline}
An error can only occur if there exists some $(m_{1c},m_{1p},m_{2c})\in\mathscr{T}$ where $(m_{1c},m_{1p})\ne(M_{1c},M_{1p})$. We divide this event into the following 4 error events:
\begin{align}
	&\mathcal{E}_{1} = \left\{\exists \, \tilde{m}_{1p}\neq M_{1p}: ( M_{1c}, \tilde{m}_{1p},M_{2c})\in\mathscr{T} \right\} \ \\
	&\mathcal{E}_{2} = \left\{ \exists \, \tilde{m}_{1c} \neq M_{1c},\tilde{m}_{1p}:( \tilde{m}_{1c}, \tilde{m}_{1p},M_{2c})\in\mathscr{T} \right\}\\
	&\mathcal{E}_{3} = \left\{\exists \,  \tilde{m}_{1p} \neq M_{1p},\tilde{m}_{2c}\neq M_{2c}: ( M_{1c}, \tilde{m}_{1p},\tilde{m}_{2c})\in\mathscr{T} \right\}\\
	&\mathcal{E}_{4} = \left\{\exists \, \tilde{m}_{1c} \neq M_{1c},\tilde{m}_{1p},\tilde{m}_{2c}\neq M_{2c}:( \tilde{m}_{1c}, \tilde{m}_{1p},\tilde{m}_{2c})\in\mathscr{T} \right\}.
	\end{align}
We will prove that the probability of each one of the error events converges to zero as long as the conditions in \eqref{eq:2} are satisfied.

To bound these error events, we state two versions of a Gaussian AVC packing lemma. The basic approach to these lemmas originates in \cite[Lemma 3]{Csiszar1988} and \cite[Lemma 1]{Csiszar}, and the proof is most similar to that of \cite[Lemma 3]{Csiszar1988}. These earlier lemmas showed that a single random codebook satisfies several desirable properties with high probability. Here, we need to show that multiple codebooks simultaneously satisfy desirable properties; thus we need a slightly more general approach.
Furthermore, we use Gaussian codewords instead of codewords uniformly distributed on the unit ball. The advantage of Gaussian codewords is that superpositions of codewords are themselves Gaussian, and we here are dealing with the summation of more than one codeword. Lemma~\ref{lem3} shows that with high probability, two superposed Gaussian codebooks yield small probability of error. While the result is stated for two codebooks for simplicity, it applies for any number of codebooks, and so it will be used to bound events $\mathcal{E}_2$, $\mathcal{E}_3$, and $\mathcal{E}_4$. Note that Lemma~\ref{lem3} requires $\Lambda<1$; i.e. the jammer's power must be less than the codeword power, which is necessary to avoid symmetrization. This requirement leads to the conditions that $S_i>J_i$ and $\alpha_i S_i+\bar\alpha_j I_i>J_i$. The latter, originating from event $\mathcal{E}_3$, constitutes the main difference between our inner bound and the Han-Kobayashi inner bound, but we have been unable to eliminate it.

Lemma~\ref{lem4} differs from Lemma~\ref{lem3} in that it focuses on just one codebook, but takes into account common randomness at the encoder and decoder. This lemma is used to bound event $\mathcal{E}_1$, because in this error event, the common message can be used as common randomness. The advantage of AVC coding with common randomness is that it is not susceptible to symmetrization. Thus, in Lemma~\ref{lem4} there is no requirement that $\Lambda<1$.

Lemmas~\ref{lem4}, \ref{lem3} and \ref{lem5} are proved in Appendix \ref{ApnB}, \ref{ApnA} and \ref{ApnC} respectively.

\begin{lemma}\label{lem4} Fix $\sigma^2$, $\Lambda \ge 0$, $N=2^{nR}$ and $K\geq n^2$. Let $\bX_i(k)$ for $i=1,\ldots,N$, $k=1,\ldots,K$ be independent zero mean Gaussian random vectors with covariance matrix $\mathbf{I}_n$. Let $R$ satisfy $R<C(1/(\Lambda+\sigma^2))$. Define 
\begin{multline}
	p_2\left(\bx_1(1),\ldots,\bx_1(K),\bx_2(1),\ldots,\bx_2(K),\ldots,
	\bx_{N}(1),\ldots,\bx_{N}(K) |\bw\right) = \\
	\frac{1}{NK}\! \sum_{i=1}^N \! \sum_{k=1}^K \!\bbP\!\left\{\exists j\!\neq\! i\!: \! \|\bx_i(k) \!+\! \bw\! +\! \bV\! -\!  \bx_j(k)\|^2\! \leq\! \|\bw \! +\!  \bV\|^2\!, \bx_i(k)\! \in\! \calT_{\eps}^{(n)}(X), \bx_j(k)\! \in\! \calT_{\eps}^{(n)}(X)\!\right\}
\end{multline}
where $\bV$ is Gaussian noise distributed as $\bV\sim\mathcal{N}(\mathbf{0},\sigma^2\mathbf{I}_n)$.
There exists $\rho>0$ such that
\begin{align}
	\lim_{n\to\infty} \bbP\left[\sup_{\bw:||\bw||^2\le n\Lambda} p_2(\bX_1(1),\ldots,\bX_1(K),\ldots, \bX_{N}(1),\ldots,\bX_{N}(K)|\bw)\le \rho\right]=0.
\end{align}
\end{lemma}

\begin{lemma}\label{lem3} Fix $\theta \in [0,1]$, $\sigma^2$, $ \Lambda\ge 0$, $N_1=2^{nR_1}$ and $N_2=2^{nR_2}$. Let $\bX_1,\ldots,\bX_{N_1} \in \mathbb{R}^n$ and $\bY_1,\ldots,\bY_{N_2} \in \mathbb{R}^n$ be independent zero mean Gaussian random vectors (codebooks) with covariance matrices $\theta \, \mathbf{I}_n$ and $\bar{\theta} \, \mathbf{I}_n$, respectively. Let $\Lambda,R_1,R_2$ satisfy

\begin{align}
\Lambda&<1,\label{eq:lambda_assumption}\\
R_1&<C\left(\frac{\theta}{\Lambda+\sigma^2}\right),\label{eq:R1_assumption}\\
R_2&<C\left(\frac{\bar\theta}{\Lambda+\sigma^2}\right),\label{eq:R2_assumption}\\
R_1+R_2&<C\left(\frac{1}{\Lambda+\sigma^2}\right).\label{eq:R12_assumption}
\end{align}
Let $\calU$ be the set of triples $(\bx,\by,\bz)\in\T(X,Y,Z)$ for some Gaussian triple $(X,Y,Z)$ where
\be
\bbE X^2=\theta,\quad \bbE Y^2=\bar\theta,\quad (X,Y,Z)\text{ are mutually independent}.
\ee
Define 
\begin{multline}
	p_1(\bx_1,\ldots,\bx_{N_1},\by_1,\ldots,\by_{N_2}|\bw) = \\
	\frac{1}{N_1N_2} \sum_{i_1=1}^{N_1}\sum_{i_2=1}^{N_2}\bbP\big\{\exists j_1\ne i_1,j_2\ne i_2:
\|\bx_{i_1}+\by_{i_2}+\bw+\bV-\bx_{j_1}-\by_{j_2}\|^2\le \|\bw+\bV\|^2,\\
(\bx_{i_1},\by_{i_2},\bw+\bV)\in\calU, (\bx_{j_1},\by_{j_2},\bx_{i_1}+\by_{i_2}+\bw+\bV-\bx_{j_1}-\by_{j_2})\in\calU\big\}
\end{multline}
where $\bV$ is Gaussian noise distributed as $\bV\sim\mathcal{N}(\mathbf{0},\sigma^2\mathbf{I}_n)$.
There exists $\rho>0$ such that
\begin{align}
	\lim_{n\to\infty} \bbP\left[\sup_{\bw:||\bw||^2\le n\Lambda} p_1(\bX_1,\ldots,\bX_{N_1},\bY_1,\ldots,\bY_{N_2}|\bw)\le\exp(-n\rho)\right]=0.\label{eq:two_codebooks}
\end{align}
\end{lemma}

Since we are dealing with more than one Gaussian codeword in this paper, we need a new version of \cite[Lemma 3]{Csiszar1988} and \cite[Lemma 1]{Csiszar} not only for Gaussian vectors, but also for multiple codebooks. It did not appear possible to use the properties derived from these lemmas on each codebook individually; instead, we must prove a new lemma establishing joint properties among more than one codebook. This new lemma (Lemma \ref{lem5}) provides the main properties that the Gaussian codebooks need to satisfy in order to bound the probability of error event $\mathcal{E}_0$ by \eqref{eq:two_codebooks_c0} and as part of the proof of Lemma \ref{lem3} by \eqref{eq:two_codebooks_c1}-\eqref{eq:two_codebooks_c3}.

\begin{lemma}\label{lem5} Fix $\theta \in [0,1]$, $N_1=2^{nR_1}$ and $N_2=2^{nR_2}$. Given any random variables $X,Y,W$, define the quantity
\be
J_{X;Y;W}(R_1,R_2)=\max\big\{0,R_1-I(X;WY),R_2-I(Y;WX),R_1+R_2-I(XY;W)-I(X;Y)\big\}.
\ee
Let $\bX_{i_1}$ and $\bY_{i_2}$ be Gaussian i.i.d. $n$-length random vectors (codebooks) independent from each other with zero mean and $\cov(\bX_{i_1}) = \theta \, \bI_n$, $\cov(\bY_{i_2}) = \bar{\theta}\, \bI_n$ where $i_1\in \{1,2,\ldots,N_1\}$ and $i_2\in \{1,2,\ldots,N_2\}$. With probability approaching 1 as $n\to\infty$, they satisfy the following, for any $\bx,\by,\bw$ where $\|\bw\|^2\le n\Lambda$ and any zero mean jointly Gaussian random vector $(X,Y,X',Y',W)$ with positive definite covariance matrices with diagonals at most $(\theta,\bar{\theta},\theta,\bar{\theta},\Lambda)$.
\begin{align}
\hspace{-1em}\frac{1}{N_1N_2} \left|\left\{(i_1,i_2): (\bx_{i_1},\by_{i_2},\bw)\notin \bigcup_{\substack{ (X,Y,W)\text{ mutually independent}: \\EX^2=\theta, EY^2=\bar\theta, EW^2\le\Lambda}} \T(X,Y,W)\right\}\right| \le \exp(-n\delta(\eps)).\label{eq:two_codebooks_c0}
\end{align}
\begin{align}
&\left|\left\{(i_1,i_2):(\bx_{i_1},\by_{i_2},\bw)\in\T(X',Y',W)\right\}\right|
\le  \exp\big\{n\big[J_{X';Y';W}(R_1,R_2)+\delta(\eps)\big]\big\}\label{eq:two_codebooks_c1}
\\
&\big|\big\{(i_1,i_2):(\bx,\by,\bx_{i_1},\by_{i_2},\bw)\in\T(X,Y,X',Y',W)\big\}\big|
\le \exp\big\{n\big[J_{X';Y';XYW}(R_1,R_2)+\delta(\eps)\big]\big\}\label{eq:two_codebooks_c2}
\\
&\frac{1}{N_1N_2}\big|\big\{(i_1,i_2):(\bx_{i_1},\by_{i_2},\bx_{j_1},\by_{j_2},\bw)\in\T(X,Y,X',Y',W)\text{ for some }j_1\ne i_1,j_2\ne i_2\big\}\big|\nonumber\\
& \hspace{30em}\le8\exp\{-n\delta(\eps)/4\}\nonumber
\\&\text{if }J_{X';Y';W}(R_1,R_2)\le I(XY;X'Y'W)-2\delta(\eps).\label{eq:two_codebooks_c3}
\end{align}
\end{lemma}

We now consider each of the five error events, beginning with $\mathcal{E}_0$. Define set $\calT^{(n,k)}_{\eps}$ as $\bigcup \, \T(X_1, \ldots, X_k)$ over all joint Gaussian distributions $X_1, \ldots, X_k$ such that $(X_1, \ldots, X_k)$ are mutually independent. For every $\eps>\eps'$, we have 
\begin{align}
	\bbP(\mathcal{E}_0)&=\bbP\{(M_{1c},M_{1p},M_{2c})\notin \mathscr{S}\}\\
	& = \bbP\left\{(\bx_{1c}(M_{1c}), \bx_{1p}(M_{1c},M_{1p}),\bx_{2c}(M_{2c}),\bY_1)\notin \bigcup \, \T(X_{1c},X_{1p},X_{2c},Y_1)\right\}\label{E0_1}\\
	& = \bbP\left\{(\bx_{1c}, \bx_{1p},\bx_{2c},h_{12}\bx_{2p}+g_1 \bw_1 + \bV_1)\notin \calT^{(n,4)}_{\eps}\right\}\label{E0_2}\\
	& \leq \bbP\left\{(\bx_{1c}, \bx_{1p},\bx_{2c},\bx_{2p},\bw_1,\bV_1)\notin \calT^{(n,6)}_{\eps}\right\}\label{E0_3}\\
	& \leq \bbP\left\{(\bx_{1c}, \bx_{1p},\bx_{2c},\bx_{2p},\bw_1)\notin \calT^{(n,5)}_{\eps'}\right\}\nonumber\\
	&\quad{}+\bbP\left\{(\bx_{1c}, \bx_{1p},\bx_{2c},\bx_{2p},\bw_1,\bV_1)\notin \calT^{(n,6)}_{\eps}\Big|\,(\bx_{1c}, \bx_{1p},\bx_{2c},\bx_{2p},\bw_1)\in \calT^{(n,5)}_{\eps'}\right\}\label{E0_4}
\end{align}
where the union in \eqref{E0_1} is exactly the one in set $\mathscr{S}$ definition \eqref{setS}. \eqref{E0_3} follows because if $(\bx_{1c}, \bx_{1p},\bx_{2c},\bx_{2p},\bw_1,\bV_1)$ is typical for independent distributions then $(\bx_{1c}, \bx_{1p},\bx_{2c},h_{12}\bx_{2p}+g_1 \bw_1 + \bV_1)$ would be typical. The probabilities in \eqref{E0_4} follows from the fact that if $(\bx_{1c}, \bx_{1p},\bx_{2c},\allowbreak \bx_{2p},\bw_1)\notin \calT^{(n,5)}_{\eps'}$ then $(\bx_{1c}, \bx_{1p},\bx_{2c},\bx_{2p},\bw_1,\bV_1)\notin\calT^{(n,6)}_{\eps} $. Finally, as $n\to\infty$ the first term in \eqref{E0_4} vanishes exponentially by using the general version of Lemma \ref{lem5}-\eqref{eq:two_codebooks_c0}, and the second term in \eqref{E0_4} tends to zero by using conditional typicality lemma (see \eqref{CondTypLem} in appendix \ref{ApnA}). Then, $\bbP(\mathcal{E}_0)$ tends to zero as $n\to\infty$.

To bound the probability of event $\mathcal{E}_1$, we apply Lemma~\ref{lem4} with the following:
\begin{itemize}
  \item $i=m_{1p}$, $j=\tilde{m}_{1p}$, $k = m_{1c}$,
	\item $\bx_i(k)= \frac{h_{11}\bx_{1p}(m_{1c},m_{1p})}{\sqrt{ (1-\gamma)\alpha_1 \sigma^2 S_1}}$,\\
  \item $\bx_j(k)= \frac{h_{11}\bx_{1p}(m_{1c},\tilde{m}_{1p})}{\sqrt{(1-\gamma)\alpha_1 \sigma^2 S_1}}$,\\
	\item $\bV = \frac{h_{12}\bx_{2p}(M_{2c},M_{2p})+\bV_1}{\sqrt{(1-\gamma)\alpha_1 \sigma^2 S_1}}$,\\
	\item  $\bw = \frac{g_1\bw_1}{\sqrt{(1-\gamma)\alpha_1 \sigma^2 S_1}}$.\\
\end{itemize}
In this case, $K=2^{nR_{1c}}\ge n^2$ for sufficiently large $n$ as long as $R_{1c}>0$. Note that event $\mathcal{E}_1$ occurs if 
\begin{multline}
	\|h_{11}\bX_{1p}(m_{1c},m_{1p})+g_1\bw_1+h_{12}\bX_{2p}(M_{2c},M_{2p})+\bV_1-h_{11}\bX_{1p}(m_{1c},\tilde{m}_{1p}) \|^2\\
	\leq \|g_1\bw_1+h_{12}\bX_{2p}(M_{2c},M_{2p})+\bV_1\|^2.
	\end{multline}
Thus, by Lemma~\ref{lem4}, if $R_{1p}<C\left(\textstyle{\frac{(1-\gamma)\alpha_1 S_1}{1+J_1+(1-\gamma)\alpha_2 I_1}}\right)=C\left(\frac{(1-\gamma)\alpha_1S_1^\prime}{1+(1-\gamma)\alpha_2 I_1^\prime}\right)$, then with high probability the codebook $\bX_{1p}$ will be such that $\mathbb{P}(\mathcal{E}_1)\to 0$ as $n\to\infty$.

We now bound event $\mathcal{E}_2$ by applying Lemma~\ref{lem3} with the following particularizations:
\begin{itemize}
  \item $i_1\!=\!m_{1c}$,~$i_2\!=\!(m_{1c},m_{1p})$,~$j_1\!=\!\tilde{m}_{1c}$,~$j_2~\!=\!~(\tilde{m}_{1c},\tilde{m}_{1p})$,
	\item $\bx_{i_1}= \frac{h_{11}\bx_{1c}(m_{1c})}{\sqrt{(1-\gamma) \sigma^2 S_1}}$, $\by_{i_2}= \frac{h_{11}\bx_{1p}(m_{1c},m_{1p})}{\sqrt{(1-\gamma) \sigma^2 S_1}}$,\\
  \item $\bx_{j_1}= \frac{h_{11}\bx_{1c}(\tilde{m}_{1c})}{\sqrt{(1-\gamma) \sigma^2 S_1}}$, $\by_{j_2}= \frac{h_{11}\bx_{1p}(\tilde{m}_{1c},\tilde{m}_{1p})}{\sqrt{(1-\gamma) \sigma^2 S_1}}$, \\
	\item $\bV = \frac{h_{12}\bx_{2p}(M_{2c},M_{2p})+\bV_1}{\sqrt{(1-\gamma) \sigma^2 S_1}}$,\\
	\item  $\bw = \frac{g_1\bw_1}{\sqrt{(1-\gamma) \sigma^2 S_1}}$.\\
\end{itemize}
Note that event $\calE_2$ occurs if
\begin{align*}
	\|h_{11}\bX_{1c}(m_{1c})\!+\! h_{11}\bX_{1p}(m_{1c},m_{1p})\!+\! \bw\!+\! \bV\!-\! h_{11}\bX_{1c}(\tilde{m}_{1c})\!-\! h_{11}\bX_{1p}(\tilde{m}_{1c},\tilde{m}_{1p}) \|^2
	\leq \|\bw\!+\! \bV\|^2.
	\end{align*}
Therefore, we can conclude by Lemma~\ref{lem3} that with high probability as $n\to \infty$, $\mathbb{P}(\mathcal{E}_{2})\to 0$ if $\frac{J_1}{S_1}<1$,
\begin{align}
	R_{1c}&<C\left(\textstyle{\frac{(1-\gamma)\bar{\alpha}_1 S_1}{1+J_1+(1-\gamma)\alpha_2 I_1}}\right)\label{R1c}\\
R_{1p}&<C\left(\textstyle{\frac{(1-\gamma)\alpha_1S_1}{1+J_1+(1-\gamma)\alpha_2 I_1}}\right)\label{R1p}\\
R_{1c}+R_{1p}&<C\left(\textstyle{\frac{(1-\gamma)S_1}{1+J_1+(1-\gamma)\alpha_2 I_1}}\right).\label{R1c_R1p}
\end{align}
Similarly, the probability of event $\calE_3$ can be bounded using Lemma~\ref{lem3} as long as we have $\alpha_1 S_1+\bar{\alpha}_2 I_1>J_1$, \eqref{R1p}, 
\begin{align}
R_{2c}<C\left(\textstyle{\frac{(1-\gamma)\bar{\alpha}_2 I_1}{1+J_1+(1-\gamma)\alpha_2 I_1}}\right)\label{R2c}\\
R_{1p}+R_{2c}<C\left(\textstyle{\frac{(1-\gamma)(\alpha_1 S_1+\bar{\alpha}_2 I_1)}{1+J_1+(1-\gamma)\alpha_2 I_1}}\right)\label{R1p_R2c}.
\end{align} 
Finally, the probability of event $\mathcal{E}_4$ may be bounded using Lemma~\ref{lem3} under the conditions $S_1 + \bar{\alpha}_2 I_1>J_1$, \eqref{R1c}, \eqref{R1p}, \eqref{R1c_R1p}, \eqref{R2c}, \eqref{R1p_R2c}, $R_{1c}+R_{2c}<C\left(\textstyle{\frac{(1-\gamma)(\bar{\alpha}_1 S_1 + \bar{\alpha}_2 I_1)}{1+J_1+(1-\gamma)\alpha_2 I_1}}\right)$ and $R_{1c}+R_{1p}+R_{2c}<C\left(\textstyle{\frac{(1-\gamma)(S_1 + \bar{\alpha}_2 I_1)}{1+J_1+(1-\gamma)\alpha_2 I_1}}\right)$. Note that in this case we use a version of Lemma~\ref{lem3} for three independent codebooks rather than two. We finally get all equations in \eqref{eq:2} as $\gamma\to 0$. 

\section{Conclusion}\label{7}

We provided inner and outer bounds for the capacity region of the Gaussian interference channel in the presence of malicious jammers. We showed that if input power is less than the jammers' power, then the capacity region is zero. Otherwise, both our inner and outer bounds correspond to existing bounds with the noise variance increased by the jammers' received power, except that our inner bound has an additional constraint compared to the classical Han-Kobayashi bound. These bounds allowed us to recover most existing results for the GIC with no jammer, with the exception of the half-bit gap, which we have been unable to prove in general. These results include the symmetric DoF of our channel as signal-to-noise ratio tends to infinity along with the interference-to-noise ratio and the jammer-to-noise ratio. To prove our achievability result, we employed three novel lemmas, including new versions of the packing lemma for multiple Gaussian codebooks. Extensions to more than two users should be straightforward. We believe that our new packing lemmas may be applicable to other network information theory problems in the presence of a malicious intruder.

\appendices

\section{Proof of Proposition \ref{Prp2}}\label{Apn0}

First, we prove that $\mathscr{C} \subseteq \mathscr{C}_{G}$. Consider rate pair $(R'_1,R'_2)\in \mathscr{C}$ meaning that there exists a $\left(2^{nR'_1},2^{nR'_2},n\right)$ code that yields an arbitrary small probability of error for any possible adversary action of two independent jammers. Now, we must show that this same code will also work for $G$ jammers. In the $G$ jammer model, let $\bw_1,\ldots,\bw_G$ be any jamming signals. We may define equivalent jamming signals for the model with two independent jammers as
\[\bw'_1=\frac{g_{11} \bw_1 +g_{12} \bw_2 +\ldots +g_{1G} \bw_G}{g_1}\]
and
\[\bw'_2=\frac{g_{21} \bw_1 +g_{22} \bw_2 +\ldots +g_{2G} \bw_G}{g_2}.\]
Note that the received signal in the $G$ jammer model is identical to that in the 2 jammer model with jamming signals $\bw_1',\bw_2'$. Moreover, in order to show that $\bw'_1$ and $\bw'_2$ satisfy power constraints, we have 
\begin{align}
	\|\bw'_i\|^2&= \frac{\|g_{i1} \bw_1 +g_{i2} \bw_2 +\ldots +g_{iG} \bw_G\|^2}{|g_i|^2}\\
	&\leq \frac{\left(|g_{i1}| \|\bw_1\| +|g_{i2}| \|\bw_2\| +\ldots +|g_{iG}| \|\bw_G\|\right)^2}{|g_i|^2}\\
	&\leq \frac{\left(|g_{i1}| \sqrt{n\Lambda} +|g_{i2}| \sqrt{n\Lambda} +\ldots +|g_{iG}| \sqrt{n\Lambda}\right)^2}{|g_i|^2}\\
	&\leq n\Lambda \frac{\left(|g_{i1}| +|g_{i2}|+\ldots +|g_{iG}|\right)^2}{|g_i|^2}\\
	&= n\Lambda\label{last}
\end{align}
where $i=1,2$, and \ref{last} follows from the assumption in the proposition. Therefore, the probability of error under the $G$ jammer model is at most that of the 2 jammer model.

Now, we prove $\mathscr{C}_{G} \subseteq \mathscr{C}$. Let rate $(R_1,R_2)\in \mathscr{C}_{G}$. Therefore, there exists a sequence of $\left(2^{nR_1},2^{nR_2},n\right)$ code that has arbitrary small probability of error $P_{e}^{G}$ for any possible adversary actions with $G$ jammers. Now, if we use this code for two-jammer scenario Fig.~\ref{fig:Fig1Label}, the probability of error at decoder $i=1,2$ is given as
\begin{align}
	P_{ei}&=\max_{\bw'_i} \bbP(\hat{M}_i \neq M_i )\\
	&=\max_{\bw_1,\bw_2,\ldots,\bw_G} \bbP(\hat{M}_i \neq M_i )\label{MAXoverW}
\end{align}
where the last equality follows because of the same power constraints $\|\bw_j\|^2\leq n\Lambda$ and $\|\bw'_i\|^2 \leq n\Lambda$ \eqref{last} for $j=1,2,\ldots,G$ and $i=1,2$ in both models meaning that the set of received jammer signals $g_{i1} \bw_1 +g_{i2} \bw_2 +\ldots +g_{iG} \bw_G$ or $g_{i} \bw'_i$ are identical at each decoder. Note that we have the assumption $|g_{i1}| +|g_{i2}|+\ldots +|g_{iG}|=|g_i|$ for $i=1,2$.
By the equivalent expression for $P_{ei}$ in \eqref{MAXoverW}, the probability of error in the $G$ jammer model can be lower bounded by
\begin{align}
	P_{e}^{G} \geq \max \{P_{e1},P_{e2}\}\label{maxError}.
\end{align}
In addition, the overall probability of error is upper bounded by 
	\begin{align}
P_{e}^{(n)} \leq P_{e1}+P_{e2}.\label{sumError}		
	\end{align}
Since $P_{e}^{G}\to 0$, from \eqref{maxError} both $P_{e1}$ and $P_{e2}$ must tend to zero, too. Therefore, the sum in \eqref{sumError} also tends to zero, so the overall probability of error with two jammers $P_{e}^{(n)}$ vanishes as $n\to\infty$.

\section{Proof of Lemma~\ref{lem4}}\label{ApnB}

We prove this lemma using a random code reduction, as in \cite[Lemma 12.8]{CsiszarKorner}. We first show that a Gaussian codebook independent of the jammer's signal achieves small probability of error, and then we show that a finite number of deterministic codebooks achieve essentially the same probability.

Let $\bX_1,\ldots,\bX_N$ be Gaussian random vectors with zero mean and covariance $ \bI_n$. We will prove that, for any $i\in[N]$ and any $\bw$ such that $\|\bw\|^2\le n\Lambda$
\begin{equation}\label{eq:gaussian_codebook}
\bbP\big\{\exists j\ne i: \|\bX_i+\bw+\bV-\bX_j\|^2\le \|\bw+\bV\|^2\big\}\to 0
\end{equation}
as $n\to\infty$, where $\bV\sim\calN(0,\sigma^2\bI_n)$. To prove this, we adopt the basic approach of \cite{Lapidoth1996}. In particular, let $\bZ=\bw+\bV$, and let $U$ be a unitary matrix that maps $\bZ$ to $(\|\bZ\|,0,\ldots,0)$. Then we may write
\begin{align}
\bbP\big\{\exists j\ne i: \|\bX_i+\bZ-\bX_j\|^2\le \|\bZ\|^2\big\}&=\bbP\big\{\exists j\ne i: \|U \bX_i+U\bZ-U\bX_j\|^2\le \|U\bZ\|^2\big\}
\\&= \bbP\big\{\exists j\ne i: \|\bX_i+U\bZ-\bX_j\|^2 \le \|U\bZ\|^2\big\}\label{eq:unitary}
\end{align}
where \eqref{eq:unitary} follows from the spherical symmetry of the codebook distribution. Now if we define, for any $\Sigma>0$,
\begin{align}
	e(\Sigma)=\bbP\{\exists j\ne i: \|\bX_i+(\sqrt{n\Sigma},0,\ldots,0)-\bX_j\|^2\le n\Sigma\}
\end{align}
then the probability in \eqref{eq:unitary} may be written as $\bbE e(\|\bZ\|^2)$. Note that for any $\delta$,
\begin{align}
	\lim_{n\to\infty} \bbP\big\{\big| \|\bZ\|^2-n(\sigma^2+\Lambda)\big|>n\delta\}\to 0.
\end{align}
Moreover, $e(\Sigma)$ is non-decreasing in $\Sigma$. Thus, for any $\delta>0$, if we let $\tilde\bV\sim\calN(0,\sigma^2+\Lambda+\delta)$, for sufficiently large $n$ we have
$
\bbE e(\|\bZ\|^2)\le \bbE e(\|\tilde\bV\|^2).
$
Now, $\bbE e(\|\tilde\bV\|^2)$ is simply the probability of error for a Gaussian channel with noise variance $\sigma^2+\Lambda+\delta$. Since Gaussian codebooks achieve capacity for Gaussian channels with minimum distance decoding, this quantity vanishes with $n$ as long as
\begin{align}
	R<C\left(\frac{1}{\sigma^2+\Lambda+\delta}\right)
\end{align}
which holds for small enough $\delta$ by the assumption that $R<C(\frac{1}{\sigma^2+\Lambda})$. This proves \eqref{eq:gaussian_codebook}.

Now let $\bX_i(k)$ for $i=1,\ldots,N$ and $k=1,\ldots,K$ be independent Gaussian vectors with zero mean and covariance $ \bI_n$. For any $i\in[N]$, $k\in[K]$, and $\bw$ such that $\|\bw\|^2\le n\Lambda$, let
\begin{multline}
\calE(k,i,\bw)=\bbP\Big\{\exists j\ne i:\|\bX_i(k)+\bw+\bV-\bX_j(k)\|^2\le \|\bw+\bV\|^2, 
\\ \bX_i(k)\in\calT_{\eps'}^{(n)}, \bX_j(k)\in\calT_{\eps'}^{(n)}\Big| \bX_1(k),\ldots,\bX_N(k)\Big\}.
\end{multline}
To prove the lemma, we need to show that, for any $\eps>0$
\begin{equation}\label{eq:probability_goal}
\lim_{n\to\infty} \bbP\left\{ \bigcup_{\bw:\|\bw\|^2\le n\Lambda}\left\{ \frac{1}{NK} \sum_{i=1}^N \sum_{k=1}^K \calE(k,i,\bw)>\eps\right\}\right\}\to 0.
\end{equation}
From \eqref{eq:gaussian_codebook}, we know that for any $\delta>0$ and sufficiently large $n$ for any $k,i,\bw$, we have $\bbE\calE(k,i,\bw)\le \delta$ (The conditions on $\bX_i$ and $\bX_j$ only decrease the probability.) Thus, for fixed $i$ and $\bw$ we have
\begin{align}
\bbP\left\{\! \frac{1}{K}\sum_{k=1}^K \calE(k,i,\bw)\!>\!\eps/2\right\}
&\!=\! \bbP\left\{2^{\sum_{k=1}^K \calE(k,i,\bw)}>2^{K\eps/2}\right\}
\\&\le 2^{-K\eps/2} \prod_{k=1}^K \bbE 2^{\calE(k,i,\bw)}\label{eq:bernstein1}
\\&\le 2^{-K\eps/2} (1+\bbE \calE(1,i,\bw))^K\label{eq:bernstein2}
\\&= 2^{-K(\eps/2-\log(1+\delta))}
\end{align}
where \eqref{eq:bernstein1} holds by Markov's inequality, and \eqref{eq:bernstein2} holds since $2^t\le 1+t$ if $t\in[0,1]$. Thus, if we let $\bw_1,\ldots,\bw_L$ be any finite set of vectors with norm at most $\sqrt{n\Lambda}$, we may apply the union bound to find
\begin{align}
	\bbP\left\{\bigcup_{\substack{l\in[L]\\ i\in[N]}} \left\{\frac{1}{K}\sum_{k=1}^K \calE(k,i,\bw_l)>\frac{\eps}{2}\right\}\right\}
	\le LN 2^{-K(\frac{\eps}{2}-\log(1+\delta))}.
\end{align}
In particular, let $\bw_1,\ldots,\bw_L$ be a $\nu$-dense subset of points in the sphere of radius $\sqrt{n\Lambda}$. There exists such a set with $L=2^{n\rho}$ for some $\rho$. Since $\calE(k,i,\bw)$ is continuous in $\bw$, for sufficiently small $\nu$, if the probability of error for all $\bw_l$ is at most $\eps/2$, then the probability of error for all $\bw$ is at most $\eps$. Thus we may bound the probability in \eqref{eq:probability_goal} by
\[
2^{n\rho} 2^{nR} 2^{-K(\eps/2-\log(1+\delta))}.
\]
As long as $\delta$ is small enough so that $\eps/2-\log(1+\delta)>0$ and $K/n\to\infty$, this probability vanishes in $n$.

\section{Proof of Lemma~\ref{lem3}}\label{ApnA}

Throughout the appendices, we use the following \emph{conditional typicality lemma} and \emph{joint typicality lemma} for Gaussian random variables; their proofs easily follow from the corresponding proofs in \cite{ElGamal} for discrete memory-less random variables.

\emph{Conditional Typicality Lemma}:
Let $(X,Y)\sim f(x,y)$. Suppose that $\bx \in \T(X)$ and $\bY \sim f(\by|\bx)=\prod_{i=1}^{n}{f_{Y|X}(y_i|x_i)}$. Then, for every $\eps >\eps'$,
\begin{align}
	\lim_{n\to \infty}\bbP\{(\bx,\bY)\in\T(X,Y)\}=1.\label{CondTypLem}
\end{align}

\emph{Joint Typicality Lemma}:
Let $(X,Y,Z)\sim f(x,y,z)$. If $(\tilde{\bx},\tilde{\by})$ is a pair of arbitrary sequences and $\tilde{\bZ}\sim \prod_{i=1}^n{f_{Z|X}(\tilde{z}_i|\tilde{x}_i)}$ then there exists $\delta(\eps)>0$ that tends to zero as $\eps \to 0$ such that
\begin{align}
	\bbP\{(\tilde{\bx},\tilde{\by},\tilde{\bZ})\in \T(X,Y,Z)\}\leq \exp(-n(I(Y;Z|X)-\delta(\eps))).
\end{align}

We apply Lemma \ref{lem5} to assume the two codebooks satisfy \eqref{eq:two_codebooks_c0}--\eqref{eq:two_codebooks_c3}. To prove \eqref{eq:two_codebooks}, first note that by \eqref{eq:two_codebooks_c0}, with high probability $(\bx_{i_1},\by_{i_2},\bw)\in \mathcal{T}_{\eps'}^{(n)}(X,Y,W)$   where $(X,Y,W)$ are mutually independent, and
\be
\bbE X^2=\theta,\quad \bbE Y^2=\bar\theta,\quad \bbE W^2\le \Lambda.
\ee
Thus, by the conditional typicality lemma, for every $\eps>\eps'$ with high probability $(\bx_{i_1},\by_{i_2},\bw,\bV)\in \T(X,Y,W,V)$ where $(X,Y,W,V)$ are mutually independent, and $\bbE V^2=\sigma^2$. This implies that $(\bx_{i_1},\by_{i_2},\bw+\bV)\in\calU$, and also that
\be
\|\bw+\bV\|^2\le n(\Lambda+\sigma^2+\eps).
\ee

We use shorthand $\vec{X}=(XYX'Y'WV)$. For $i_1,i_2,\bw$ and any Gaussian distribution on $\vec X$, define
\be
e_{\vec X}(i_1,i_2,\bw)= \bbP\Big\{(\bx_{i_1},\by_{i_2},\bx_{j_1},\by_{j_2},\bw,\bV)\in \T(\vec X)\text{ for some }j_1\ne i_1,j_2\ne i_2\Big\}.
\ee
We  need to show that for some $\delta>0$,
\be
\frac{1}{N_1N_2}\sum_{i_1,i_2} e_{\vec X}(i_1,i_2,\bw)\le \exp(-n\rho)
\ee
for all $\vec X$ where
\begin{gather}
(X,Y,W,V)\text{ are mutually independent},\\
\bbE X^2=\bbE X'^2=\theta,\quad
\bbE Y^2=\bbE Y'^2=\bar\theta,\quad
\bbE V^2=\sigma^2\\
\bbE W^2\le\Lambda,\quad
\bbE (X+Y+W+V-X'-Y')^2\le \Lambda+\sigma^2\\
(X',Y',X+Y+W+V-X'-Y')\text{ are mutually independent}.
\end{gather}
Observe that if $I(XYV;WX'Y')=0$, then we would have
\begin{align}
\Lambda+\sigma^2&\ge \bbE(X+Y+W+V-X'-Y')^2
\\&= \bbE (X+Y+V)^2+\bbE(W-X'-Y')^2\\
&\ge 1+\sigma^2.
\end{align}
But this cannot happen since $\Lambda<1$ by \eqref{eq:lambda_assumption}. Thus, there exists $\eta>0$ where
\be
I(XYV;WX'Y')\ge\eta.
\ee
Recalling that $I(XYV;W)=0$, this implies
\be\label{eq:eta_bound}
I(XYV;X'Y'|W)\ge\eta.
\ee
Also, by \eqref{eq:two_codebooks_c3}, we may restrict ourselves to distributions where
\be\label{eq:R1R2}
J_{X';Y';W}(R_1,R_2) \ge I(XY;X'Y'W)-2\delta(\eps).
\ee
We may now write, for any $(i_1,i_2)$ and any $\bw$
\begin{align}
e_{\vec X}(i_1,i_2,\bw)
&\le \sum_{(j_1,j_2):(\bx_{i_1},\by_{i_2},\bx_{j_1},\by_{j_2},\bw)\in\T} \bbP\Big\{(\bx_{i_1},\by_{i_2},\bx_{j_1},\by_{j_2},\bw,\bV)\in \T\Big\}
\\&\le \exp\Big\{n\Big[J_{X';Y';XYW}(R_1,R_2) - I(V;X'Y'|XYW)+\delta(\eps)\Big]\Big\}.\label{eq:two_codebook_ebound}
\end{align}
where in \eqref{eq:two_codebook_ebound} we have applied \eqref{eq:two_codebooks_c2}, the joint typicality lemma, and the fact that $I(V;XYW)=0$.

We consider two cases.

Case (a): $J_{X';Y';W}(R_1,R_2)=0$. Note that $J_{X';Y';XYW}(R_1,R_2)\le J_{X';Y';W}(R_1,R_2)$ so in this case we also have $J_{X';Y';XYW}(R_1,R_2)=0$. By \eqref{eq:R1R2}, $I(XY;X'Y'W)\le 2\delta(\eps)$. Thus, from \eqref{eq:eta_bound}
\begin{align}
\eta &\le I(XYV;X'Y'|W)\\
&=I(XY;X'Y'|W)+I(V;X'Y'|XYW)\\
&\le 2\delta(\eps)+I(V;X'Y'|XYW).
\end{align}
From \eqref{eq:two_codebook_ebound}, we have
\begin{align}
e_{\vec X}(i_1,i_2,\bw)&\le \exp\{n[- I(V;X'Y'|XYW)+\delta(\eps)]\}\\
&\le \exp\{n[-\eta+3\delta(\eps)]\}.
\end{align}
This vanishes exponentially fast if $\delta(\eps)$ is sufficiently small.

Case (b): $J_{X';Y';W}(R_1,R_2)>0$. This implies that (recalling that $I(X';Y')=0$)
\be
J_{X';Y';W}(R_1,R_2)=\max\{R_1-I(X';WY'),R_2-I(Y';WX'),R_1+R_2-I(X'Y';W)\}.
\ee
By \eqref{eq:R1R2}, we have
\be
-2\delta(\eps)\le \max\{R_1-I(X';WY'),R_2-I(Y';WX'),R_1+R_2-I(X'Y';W)\}-I(XY;X'Y'W).
\ee
Note that
\begin{align}
I(X';WY')+I(XY;X'Y'W)&\ge I(X';WY')+ I(XY;X'|WY')\\
&=I(X';XYWY').
\end{align}
Similarly
\begin{align}
I(Y';WX')+I(XY;X'Y'W)&\ge I(Y';XYWX'),
\\I(X'Y';W)+I(XY;X'Y'W)&\ge I(X'Y';XYW).
\end{align}
Thus
\be
-2\delta(\eps)\le\max\{R_1-I(X';XYWY'),R_2-I(Y';XYWX'),R_1+R_2-I(X'Y';XYW)\}.
\ee
Hence
\begin{multline}
J_{X';Y';XYW}(R_1,R_2)\\
\le \max\{R_1-I(X';XYWY'),R_2-I(Y';XYWX'),R_1+R_2-I(X'Y';XYW)\}+2\delta(\eps).
\end{multline}
By \eqref{eq:two_codebook_ebound}, we have
\begin{align}
&\frac{1}{n}\log e_{\vec X}(i_1,i_2,\bw)\nonumber
\\&\le J_{X';Y';XYW}(R_1,R_2)-I(V;X'Y'|XYW)+\delta(\eps)
\\&\le \max\{R_1-I(X';XYWY'),R_2-I(Y';XYWX'),R_1+R_2-I(X'Y';XYW)\}\nonumber
\\& \quad-I(V;X'Y'|XYW)+3\delta(\eps)
\\&\le \max\{R_1-I(X';XYWY'V),R_2-I(Y';XYWX'V),R_1+R_2-I(X'Y';XYWV)\}\nonumber
\\&\quad+3\delta(\eps).
\end{align}
Let $Z=X+Y+W+V-X'-Y'$. Recalling that $(X',Y',Z)$ are mutually independent and $\bbE Z^2\le\Lambda+\sigma^2$, we have
\begin{align}
I(X';XYWY'V)&\ge I(X';X+Y+W+V-Y')
\\&=I(X';X'+Z)
\\&= h(X'+Z)-h(X'+Z|X')
\\&= \frac{1}{2}\log 2\pi e(\theta+\bbE Z^2)-h(Z|X')
\\&= \frac{1}{2}\log 2\pi e(\theta+\bbE Z^2)-\frac{1}{2} \log 2\pi e\bbE Z^2
\\&= C\left(\frac{\theta}{\bbE Z^2}\right)
\\&\ge C\left(\frac{\theta}{\Lambda+\sigma^2}\right).
\end{align}
Similarly
\be
I(Y';XYSX'V)\ge C\left(\frac{\bar\theta}{\Lambda+\sigma^2}\right).
\ee
Moreover,
\begin{align}
I(X'Y';XYWV)&\ge I(X'Y';Z+X'+Y')
\\&=h(Z+X'+Y')-h(Z)
\\&=C\left(\frac{1}{\bbE Z^2}\right)
\\&\ge C\left(\frac{1}{\Lambda+\sigma^2}\right).
\end{align}
Thus
\be
e_{\vec X}(i_1,i_2,\bw)\le\exp\Big\{n\Big[{\textstyle\max\big\{R_1-C\big(\frac{\theta}{\Lambda+\sigma^2}\big),
R_2-C\big(\frac{\bar\theta}{\Lambda+\sigma^2}\big), 
R_1+R_2-C\big(\frac{1}{\Lambda+\sigma^2}\big)\big\}+3\delta(\eps)}\Big]\Big\}.
\ee
Therefore, $e_{\vec X}(i_1,i_2,\bw)$ is exponentially vanishing if $\delta(\eps)$ is sufficiently small and \eqref{eq:R1_assumption}--\eqref{eq:R12_assumption} hold.

\section{Proof of Lemma~\ref{lem5}}\label{ApnC}

Since we frequently use \cite[Lemma A1]{Csiszar1988} throughout this appendix, we provide the statement of this lemma here.

\begin{lemma}\label{LemA1}Let $\bZ_1,\ldots,\bZ_N$ be arbitrary random variables, and let $f_i(\bZ_1,\ldots,\bZ_i)$ be arbitrary with $0\leq f_i \leq 1$, $i=1,\ldots,N$. Then the condition 
\begin{align}
	\bbE\left[ f_i(\bZ_1,\ldots,\bZ_i)|\bZ_1,\ldots,\bZ_{i-1}\right]\leq a \text{ a.s.,} \quad{} i=1,\ldots, N,
\end{align}
implies that 
\begin{align}
	\bbP \left\{ \frac{1}{N} \sum_{i=1}^N f_i(\bZ_1,\ldots,\bZ_i)>t\right\}\leq \exp\{-N(t-a \log e)\}.
\end{align}
\end{lemma}

Let $\bX_{i_1}$ and $\bY_{i_2}$ be Gaussian i.i.d. $n$-length random vectors (codebooks) independent from each other with $\var(X) = \theta$ and $\var(Y) = \bar{\theta}$. Fix $\bx\in \calT_\eps^{(n)}(X), \by\in \calT_\eps^{(n)}(Y), \bw \in \scS^n$ and a covariance matrix $\cov(X,Y,X',Y',W)\in\calV^{5\times 5}$ such that $\scS^n$ is a $\nu$-dense subset of $\bbR^n$ for $\bw$ such that $||\bw||^2\le n\Lambda$, and $\calV^{5\times 5}$ is a $\nu$-dense subset of $\bbR^{5x5}$ for positive definite covariance matrices with diagonals at most $(\theta,\bar{\theta},\theta,\bar{\theta},\Lambda)$.

Let
\begin{align}
	A_{\eps}^{n}(X,W)=\bigcup_{\substack{X,W \text{ independent}\\ \bbE X^2 = \theta,\bbE W^2 \leq \Lambda}}\calT_\eps^{(n)}(X,W)
\end{align}
and 
\begin{align}
	A_{\eps}^{n}(X,Y,W)=\bigcup_{\substack{ (X,Y,W) \text{ mutually independent } \\EX^2=\theta, EY^2=\bar\theta, EW^2\le\Lambda}}\T(X,Y,W).
\end{align}
To prove \eqref{eq:two_codebooks_c0}, first define $h_{i_1}$ as a function of $\bX_1,\ldots, \bX_{i_1}$ as follows: 
\begin{align}
	h_{i_1}(\bX_1,\ldots, \bX_{i_1})=\begin{cases}
	1, & \text{ if } (\bX_{i_1},\bw) \notin A_{\eps}^{n}(X,W)\\
	0, & \text{ otherwise }.
	\end{cases}
\end{align}
Then the expected value of $h_{i_1}$ is given as 
\begin{align}
\bbE[h_{i_1}(\bX_1,\ldots, \bX_{i_1})|\bX_1,\ldots, \bX_{i_1-1}] &= \bbE \left[\mathbbm{1}\left((\bX_{i_1},\bw) \notin A_{\eps}^{n}(X,W) \right)\right] \\
	& = \bbP \left\{(\bX_{i_1},\bw) \notin A_{\eps}^{n}(X,W)\right\} \label{unionXW}\\
	& \leq \bbP \left\{\frac{1}{n}\|\bw\|^2 \ge \Lambda+\epsilon \right\}+\bbP \left\{\frac{1}{n}|\langle \bX_{i_1},\bw\rangle| \ge \epsilon\right\} \nonumber\\
	& \quad{}+\bbP \left\{|\frac{1}{n}\|\bX_{i_1}\|^2-\theta| \ge \epsilon \right\}\label{assume2}\\
	& \leq \exp (-n r_1(\eps))\label{LDT}
\end{align}
where \eqref{assume2} follows since the union in $A_{\eps}^{n}(X,W)$ is over independent $X,W$ such that $\bbE X^2 \!=~\theta, \allowbreak \bbE W^2 \leq \Lambda$, and directly from the definition of typical set in \eqref{TypicalSet} we obtain that it only suffices to find the probability of those $(\bX_{i_1},\bw)$ that simultaneously do not satisfy the conditions of the union and the typical set definition's inequalities. The upper bound in \eqref{LDT} follows since the first term in \eqref{assume2} is equal to zero (assumption $\|\bw\|^2\leq n\Lambda$), and the other terms are exponentially vanishing by using the large deviation theory for Gaussian distributions $X$ with positive function $r_1(\eps)$.

Now, using Lemma \ref{LemA1}, we have 
\begin{align}
	\bbP\bigg\{ \frac{1}{N_1} \left|\left\{i_1:  (\bX_{i_1},\bw)\notin A_{\eps}^{n}(X,W)\right\}\right| &> \exp(-n\delta_1(\eps))\bigg\}\nonumber\\
	& \leq \exp\{-N_1[\exp(-n\delta_1(\eps))-\exp (-n r_1(\eps))\log e ]\} \\
	& \leq \exp(-\exp(n\rho_1(\eps))).
	\end{align}
where the last inequality follows as long as $\delta_1(\eps)< r_1(\eps)$ for some $\rho_1(\eps)>0$.
Thus, the probability vanishes doubly exponentially as $n\to\infty$, and with high probability we have
\begin{align}
	\frac{1}{N_1}\left|\left\{i_1: (\bx_{i_1},\bw)\notin A_{\eps}^{n}(X,W)\right\}\right| \le \exp(-n\delta_1(\eps)).\label{i1eq1}
\end{align}

Fix $\bx_{i_1}$, and for any $i_2$ define $\tilde{h}_{i_2}$ as
	\begin{align}
		\tilde{h}_{i_2}(\bY_1,\ldots, \bY_{i_2}) = \frac{1}{N_1}\sum_{i_1: (\bx_{i_1},\bw)\in \, A_{\eps}^{n}(X,W)} \mathbbm{1}\left((\bx_{i_1},\bY_{i_2},\bw) \notin A_{\eps}^{n}(X,Y,W)\right).
\end{align}
	The expected value of $\tilde{h}_{i_2}$ can be written as
\begin{align}
\bbE\left[\tilde{h}_{i_2}(\bY_1,\ldots, \bY_{i_2})| \bY_1,\ldots, \bY_{i_2-1}\right]& = \frac{1}{N_1}\sum_{i_1: (\bx_{i_1},\bw)\in \, A_{\eps}^{n}(X,W)} \bbP\left\{(\bx_{i_1},\bY_{i_2},\bw)\notin A_{\eps}^{n}(X,Y,W) \right\}\nonumber\\
	&\leq \bbP \left\{\frac{1}{n}\|\bw\|^2 \ge \Lambda+\epsilon \right\}+\bbP \left\{\frac{1}{n}|\langle \bx_{i_1},\bw\rangle| \ge \epsilon\right\}\nonumber\\
	&\quad{}+\bbP \left\{\left|\frac{1}{n}\|\bx_{i_1}\|^2-\theta\right| \ge \epsilon \right\}+\bbP \left\{\frac{1}{n}|\langle \bY_{i_2},\bw\rangle| \ge \epsilon\right\}\nonumber\\
	&\quad{}+\bbP \left\{\frac{1}{n}\left|\langle \bY_{i_2},\bx_{i_1}\rangle \right| \ge \epsilon \right\}+\bbP \left\{\left|\frac{1}{n}\|\bY_{i_2}\|^2-\bar{\theta}\right| \ge \epsilon \right\}\label{Eh_2}\\
	& \leq \exp(-nr_2(\eps))\label{Eh_3}
\end{align}
where \eqref{Eh_2} follows directly from the definition of typical set and the union's conditions, and \eqref{Eh_3} follows since the first three terms in \eqref{Eh_2} are equal to $0$ due to the assumptions and the other terms in \eqref{Eh_2} vanish exponentially by the large deviation theory for Gaussian distributions $Y$ with positive function $r_2(\eps)$

Using Lemma \ref{LemA1}, we have
\begin{align}
	&\bbP\left\{ \frac{1}{N_1N_2} \sum_{i_1: (\bx_{i_1},\bw)\in \, A_{\eps}^{n}(X,W)}\left|\left\{i_2: (\bx_{i_1},\bY_{i_2},\bw)\notin A_{\eps}^{n}(X,Y,W)\right\}\right|> \exp(-n\delta_2(\eps))\right\} \nonumber\\
	&\quad{} \leq \exp\{-N_2(\exp(-n\delta_2(\eps)-\exp (-nr_2(\eps) )\log e ))\}\\
	&\quad{}\leq \exp(-\exp(n\rho_2(\eps)))\label{probi2}
\end{align}
 where \eqref{probi2} follows if $\delta_2(\eps)<r_2(\eps)$ for some $\rho_2(\eps)>0$. Therefore, with probability approaching 1,  as $n\to\infty$ we have 
\begin{align}
	\frac{1}{N_1N_2} \sum_{i_1: (\bx_{i_1},\bw)\in \, A_{\eps}^{n}(X,W)}
	\quad{}\left|\left\{i_2: (\bx_{i_1},\by_{i_2},\bw)\notin A_{\eps}^{n}(X,Y,W)\right\}\right|\leq \exp(-n\delta_2(\eps)).\label{i2eq1}
\end{align}

Eventually, we easily use \eqref{i1eq1} and \eqref{i2eq1} to bound the fraction in \eqref{eq:two_codebooks_c0} as follows:
\begin{align}
	\frac{1}{N_1N_2} |\{(i_1,i_2): &(\bx_{i_1},\by_{i_2},\bw)\notin  A_{\eps}^{n}(X,Y,W)\}|\nonumber\\
	&\leq \frac{1}{N_1N_2} \sum_{i_1: (\bx_{i_1},\bw)\notin \, A_{\eps}^{n}(X,W)}\left|\left\{i_2: (\bx_{i_1},\by_{i_2},\bw)\notin A_{\eps}^{n}(X,Y,W) \right\}\right| \nonumber\\
	&\quad{}+\frac{1}{N_1N_2} \sum_{i_1: (\bx_{i_1},\bw)\in \, A_{\eps}^{n}(X,W)}\left|\left\{i_2: (\bx_{i_1},\by_{i_2},\bw)\notin A_{\eps}^{n}(X,Y,W) \right\}\right|\\
& \leq \exp(-n\delta_1(\eps))+\exp(-n\delta_2(\eps))\label{eq14proof}\\
& \leq \exp(-n\delta(\eps)).
\end{align}
This proves \eqref{eq:two_codebooks_c0}.

Next, define function $f_{i_1}$ as follows:
\begin{align}
	f_{i_1}(\bX_1,\ldots, \bX_{i_1})=\begin{cases}
	1, & \text{ if } (\bX_{i_1},\bw) \in \calT_\eps^{(n)}(X',W)\\
	0, & \text{ otherwise }.
	\end{cases}
\end{align}
Then, by joint typicality lemma we get
\begin{align}
	\bbE[f_{i_1}(\bX_1,\ldots, \bX_{i_1})|\bX_1,\ldots, \bX_{{i_1}-1}] &= \bbE \left[\mathbbm{1}\left((\bX_{i_1},\bw) \in \calT_\eps^{(n)}(X',W) \right)\right] \\
	& = \bbP \left((\bX_{i_1},\bw) \in \calT_\eps^{(n)}(X',W)\right) \\
	& \leq \exp (-nI(X',W)+n\delta(\eps)).
\end{align}
Thus, using Lemma \ref{LemA1}, we have
\begin{align}\label{LemmaA1result1}
	&\hspace{-0.8em}\bbP \left\{\left|\left\{i_1: (\bX_{i_1},\bw) \in\calT_\eps^{(n)}(X',W) \right\}\right|>\exp\left(n|R_1-I(X',W)|^+ \!+ n2\delta(\eps)\right)\right\} \nonumber\\
	&\hspace{-0.8em}\leq \exp\left\{-\exp\left(n|R_1-I(X',W)|^+ \!+ n2\delta(\eps)\right)\!+ \!\exp(-n I(X',W)+n\delta(\eps)+n R_1)\log e\right\}.
\end{align}
If $R_1>I(X',W)$, \eqref{LemmaA1result1} becomes less than doubly exponentially function $\exp(-\exp(n\sigma(\eps)))$ where $\sigma(\eps)>0$ since for large enough $n$ we obtain $\exp(n\delta(\eps))>\log e$. Now, if $R_1\leq I(X',W)$ then \eqref{LemmaA1result1} is less than
\[\exp\{-\exp(n2\delta(\eps))+\exp(nR_1+n\delta(\eps)-n I(X',W))\log e\}\leq \exp(-\exp(n\sigma(\eps)))\]
where $\sigma(\eps)>0$. In both cases, this doubly decreasing exponential function vanishes as $n\to \infty$. Hence, with high probability, we have
\be
\big|\big\{i_1:(\bx_{i_1},\bw)\in\T(X',W)\big\}\big|\le \exp\big\{n\big[|R_1-I(X';W)|^++\delta(\eps)\big]\big\}.\label{(x,s)}
\ee

For any $i_2$, 
\begin{align}
&\bbP\big\{(\bx_{i_1},\bY_{i_2},\bw)\in\T(X',Y',W)\text{ for some }i_1\big\}\nonumber \\
&\qquad{}\le \sum_{i_1:(\bx_{i_1},\bw)\in \calT_\eps^{(n)}(X',W)} \bbP\big\{(\bx_{i_1},\bY_{i_2},\bw)\in\calT_\eps^{(n)}(X',Y',W)\big\}\label{fixi1}\\
&\qquad{} \le \big|\big\{i_1: (\bx_{i_1},\bw)\in \calT_\eps^{(n)}(X',W)\big\}\big| \max_{\hat{\bx}} \bbP\big\{(\hat{\bx},\bY_{i_2},\bw)\in\calT_\eps^{(n)}(X',Y',W)\big\}\\
&\qquad{}\le \exp\big\{n\big(|R_1-I(X';W)|^+-I(Y';X'W)+\delta(\eps)\big)\big\}\label{Typical1},
\end{align}
where \eqref{fixi1} follows since if $(\bx_{i_1},\bY_{i_2},\bw)$ is typical then $(\bx_{i_1},\bw)$ is always typical and \eqref{Typical1} follows from \eqref{(x,s)} and joint typicality lemma. 
Thus, applying Lemma \ref{LemA1} in the same way that we used it to get \eqref{(x,s)}, with high probability we attain 
\begin{multline}
\big|\big\{i_2:(\bx_{i_1},\by_{i_2},\bw)\in\T(X',Y',W)\text{ for some }i_1\big\}\big|\\
\le \exp\Big\{n\Big[\big|R_2+|R_1-I(X';W)|^+-I(Y';X'W)\big|^++\delta(\eps)\Big]\Big\}\label{Noi2}.
\end{multline}
Since $(\bx_{i_1},\by_{i_2},\bw)\in\calT_\eps^{(n)}(X',Y',W)$ implies $(\by_{i_2},\bw)\in\calT_\eps^{(n)}(Y',W)$, we have the simpler bound
\begin{align}
\big|\big\{i_2:(\bx_{i_1},\by_{i_2},\bw)\in\T(X',Y',W)\text{ for some }i_1\big\}\big|
&\le \big|\big\{i_2:(\by_{i_2},\bw)\in \T(Y',W)\big\}\big|
\\&\le \exp\big\{n\big[|R_2-I(Y';W)|^++\delta(\eps)\big]\big\},\label{eq:simple_i2_bound}
\end{align}
where \eqref{eq:simple_i2_bound} is similar to \eqref{(x,s)}.
Moreover, if we replace vector $\bw$ by $(\by_{i_2},\bw)$ in \eqref{(x,s)}, then for any $i_2$ we get 
\be
\big|\big\{i_1:(\bx_{i_1},\by_{i_2},\bw)\in\T(X',Y',W)\big\}\big|
\le \exp\big\{n\big[|R_1-I(X';Y'W)|^++\delta(\eps)\big]\big\}.\label{Noi1}
\ee

Therefore,
\begin{align}
&\big|\big\{(i_1,i_2):(\bx_{i_1},\by_{i_2},\bw)\in\T(X',Y',W)\big\}\big|\nonumber
\\&\le \sum_{i_2: (\bx_{i_1},\by_{i_2},\bw)\in\T(X',Y',W)\text{ for some }i_1}
\ \big|\{i_1:(\bx_{i_1},\by_{i_2},\bw)\in\T(X',Y',W)\big\}\big|
\\&\le\exp\Big\{n\Big[\big|R_2+|R_1-I(X';W)|^+-I(Y';X'W)\big|^+ + |R_1-I(X';Y'W)|^++\delta(\eps)\Big]\Big\}\label{(i1,i2)}
\end{align}
where \eqref{(i1,i2)} follows from \eqref{Noi2} and \eqref{Noi1}. 
If $R_1\le I(X';Y'W)$, then we have
\begin{align}
&\big|\big\{(i_1,i_2):(\bx_{i_1},\by_{i_2},\bw)\in\T(X',Y',W)\big\}\big|\nonumber
\\&\le \exp\Big\{n\Big[\big|R_2+|R_1-I(X';W)|^+-I(Y';X'W)\big|^++\delta(\eps)\Big]\Big\}
\\&=\exp\Big\{n\Big[\max\big\{0,R_2-I(Y';X'W),R_1+R_2-I(X'Y';W)-I(X';Y')\big\}+\delta(\eps)\Big]\Big\}
\\&= \exp\big\{n\big[J_{X';Y';W}(R_1,R_2)+\delta(\eps)\big]\big\}.
\end{align}
Using \eqref{eq:simple_i2_bound} and \eqref{Noi1}, we alternatively bound
\begin{multline}
\big|\big\{(i_1,i_2):(\bx_{i_1},\by_{i_2},\bw)\in\T(X',Y',W)\big\}\big|\\
\le \exp\big\{n\big[|R_2-I(Y';W)|^+ +|R_1-I(X';Y'W)|^++\delta(\eps)\big]\big\}.
\end{multline}
In particular, if $R_1>I(X';Y'W)$ then we have
\begin{align}
&\big|\big\{(i_1,i_2):(\bx_{i_1},\by_{i_2},\bw)\in\T(X',Y',W)\big\}\big|\nonumber
\\&\quad{}\le \exp\big\{n\big[|R_2-I(Y';W)|^+ +R_1-I(X';W|Y')+\delta(\eps)\big]\big\}
\\&\quad{}=\exp\Big\{n\Big[\max\big\{R_1-I(X';W|Y'),R_1+R_2-I(X'Y';W)\big\}+\delta(\eps)\Big]\Big\}
\\&\quad{}=\exp\big\{n\big[J_{X';Y';W}(R_1,R_2)+\delta(\eps)\big]\big\}.
\end{align}
This proves \eqref{eq:two_codebooks_c1}. An identical calculation with $(X,Y,W)$ in place of $W$ gives \eqref{eq:two_codebooks_c2}.
We indeed use Lemma \ref{LemA1} to prove that the complement of two events \eqref{eq:two_codebooks_c1} and \eqref{eq:two_codebooks_c2} happen with decreasingly doubly exponential probability for sufficiently large $n$ as follows:
\begin{multline}
\bbP\big\{\big|\big\{(i_1,i_2):(\bx_{i_1},\by_{i_2},\bw)\in\T(X',Y',W)\big\}\big|
>  \exp\big\{n\big[J_{X';Y';W}(R_1,R_2)+\delta(\eps)\big]\big\}\big\} \\ < \exp(-\exp(n\sigma(\eps))), \label{eq:doubly_exp_c1}
\end{multline}
\begin{multline}
\bbP\big\{\big|\big\{(i_1,i_2):(\bx,\by,\bx_{i_1},\by_{i_2},\bw)\in\T(X,Y,X',Y',W)\big\}\big|\\ 
> \exp\big\{n\big[J_{X';Y';XYW}(R_1,R_2)+\delta(\eps)\big]\big\}\big\} <\exp(-\exp(n\sigma(\eps)))\label{eq:doubly_exp_c2}.
\end{multline}

Now, in order to prove \eqref{eq:two_codebooks_c3}, first let 
\begin{align}
	A_{(i_1,i_2)} = \big\{(j_1,j_2): j_1<i_1,j_2\neq i_2, (\bx_{j_1},\bY_{j_2},\bw)\in \T (X',Y',W) \big\},\\
	\tilde{A}_{(i_1,i_2)} = \begin{cases}
	A_{(i_1,i_2)}, & \text{ if } |A_{(i_1,i_2)}|\leq \exp\left\{n\left(J_{X';Y';W}(R_1,R_2)+\delta(\eps)\right)\right\}\\
	\emptyset, & \text{ otherwise, }
	\end{cases}\label{A_size}
	\end{align}
	where $\tilde{A}_{(i_1,i_2)}$ is defined for fixed value of $\bx_1,\ldots,\bx_{i_1-1}$ and random $\bY_1,\ldots,\bY_{j_2}$, i.e. $\tilde{A}_{(i_1,i_2)}$ is a random set.
Define
\begin{align}
	g_{i_1}(\bx_1,\ldots,\bx_{i_1}) = \bbP\left\{(\bx_{i_1},\bY_{i_2},\bx_{j_1},\bY_{j_2},\bw)\!\in\! \T (X,Y,X',Y',W) \text{ for some } (j_1,j_2) \!\in\! \tilde{A}_{(i_1,i_2)}\right\}
\end{align}
and
\begin{align}
	\tilde{g}_{i_1}(\bx_1,\ldots,\bx_{i_1}) = \begin{cases}
	1, & \text{ if } g_{i_1}(\bx_1,\ldots,\bx_{i_1})>\exp(-n\delta(\eps)/2)\\
	\emptyset, & \text{ otherwise. }
	\end{cases}
\end{align}
It is notable that since by \eqref{eq:two_codebooks_c1} 
\begin{align}
	\bbP\left\{ \big| \big\{(j_1,j_2): (\bx_{j_1},\by_{j_2},\bw)\in \T (X',Y',W) \big\}\big|>\exp\{n(J_{X';Y';W}(R_1,R_2)+\delta(\eps))\}\right\}
\end{align}
tends to zero as $n$ grows, the probability that $\tilde{A}_{(i_1,i_2)}\ne A_{(i_1,i_2)}$ for some $i_1, i_2$ vanishes as $n\to\infty$. 

Finding the expected values of $g_{i_1}$ and $\tilde{g}_{i_1}$, we have
\begin{align}
	&\bbE [ g_{i_1}(\bX_1,\ldots,\bX_{i_1})|\bX_1=\bx_1,\ldots,\bX_{i_1-1}=\bx_{i_1-1}]\nonumber\\
	&\quad{}= \bbP\Big\{(\bX_{i_1},\bY_{i_2},\bx_{j_1},\bY_{j_2},\bw)\in \T (X,Y,X',Y',W)\text{ for some }(j_1,j_2) \in \tilde{A}_{(i_1,i_2)}\Big\}\\
	&\quad{}\leq \sum_{(j_1,j_2) \in \tilde{A}_{(i_1,i_2)}}\bbP\left\{(\bX_{i_1},\bY_{i_2},\bx_{j_1},\bY_{j_2},\bw)\in \T (X,Y,X',Y',W)\right\}\\
	&\quad{}\leq \exp ( nJ_{X';Y';W}(R_1,R_2)+\delta(\eps)) \max_{\hat{\bx},\hat{\by}}\bbP\left\{(\bX_{i_1},\bY_{i_2},\hat{\bx},\hat{\by},\bw)\in \T (X,Y,X',Y',W)\right\}\label{from_definition}\\
	&\quad{} \leq\exp\{-n(-J_{X';Y';W}(R_1,R_2)+I(XY;X'Y'W)-\delta(\eps))\}\label{joint_typicality1}\\
	&\quad{} \leq \exp(-n\delta(\eps))\label{from_condition}
\end{align} 
where \eqref{from_definition} follows since by \eqref{A_size} the size of $\tilde{A}_{(i_1,i_2)}$ is almost surely less than $\exp ( n J_{X';Y';W}(R_1,\allowbreak R_2) + \delta(\eps))$, \eqref{joint_typicality1} follows by joint typicality lemma, and \eqref{from_condition} follows by the condition in \eqref{eq:two_codebooks_c3}. Moreover, by Markov's inequality we have 
\begin{align}
	\bbE [ \tilde{g}_{i_1}(\bX_1,\ldots,\bX_{i_1})|\bX_1,\ldots,\bX_{i_1-1}] &= \bbP\{g_{i_1}(\bX_1,\ldots,\bX_{i_1})>\exp(-n\delta(\eps)/2)|\bX_1,\ldots,\bX_{i_1-1}\}\nonumber \\
	& \leq \frac{\bbE[g_{i_1}(\bX_1,\ldots,\bX_{i_1})|\bX_1,\ldots,\bX_{i_1-1}]}{\exp(-n\delta(\eps)/2)} \\
	& \leq \exp(-n\delta(\eps) + n\delta(\eps)/2)\\
	& = \exp(-n\delta(\eps)/2).
\end{align}

Therefore, using Lemma \ref{LemA1}, we have 
\begin{align}
	&\bbP\left\{\frac{1}{N_1}\sum_{i_1=1}^{N_1} \tilde{g}_{i_1}(\bx_1,\ldots,\bx_{i_1})>\exp(-n\delta(\eps)/4)\right\}\nonumber\\ 
	&\quad{}= \bbP\left\{\frac{1}{N_1}|\{i_1: g_{i_1}(\bx_1,\ldots,\bx_{i_1})>\exp(-n\delta(\eps)/2)\}|>\exp(-n\delta(\eps)/4)\right\}\\ 
	&\quad{} \leq \exp\{-\exp(nR_1)(\exp(-n\delta(\eps)/4)-\exp(-n\delta(\eps)/2)\log e)\}\\
	&\quad{} \leq \exp(-\exp(n\sigma(\eps))).
\end{align}
for $\sigma(\eps)>0$ i.e. with high probability 
\begin{align}
	\frac{1}{N_1}\sum_{i_1} \tilde{g}_{i_1}(\bx_1,\ldots,\bx_{i_1}) &= \frac{1}{N_1}|\{i_1: g_{i_1}(\bx_1,\ldots,\bx_{i_1})>\exp(-n\delta(\eps)/2)\}|\\ 
	& \leq \exp(-n\delta(\eps)/4)\label{g1less}.
\end{align}
 
Let 
\begin{align}
	f_{(i_1,i_2)}(\by_1,\ldots,\by_{i_2})
	= \begin{cases}
	1, & \text{ if } (\bx_{i_1}, \by_{i_2},\bx_{j_1},\by_{j_2},\bw) \in \T(X,Y,X',Y',W),  \;\;\\
	&\text{ for some }(j_1,j_2)\in \tilde{A}_{(i_1,i_2)} \text{ and }j_2<i_2\\
	0, & \text{ otherwise.}
	\end{cases}
\end{align}
Now, fix an $i_1$ such that $g_{i_1}(\bx_1,\ldots,\bx_{i_1})\leq \exp(-n\delta(\eps)/2)$. Therefore, we have
\begin{align}
	&\bbE\left[f_{(i_1,i_2)}(\bY_1,\ldots,\bY_{i_2})|\bY_1,\ldots,\bY_{i_2-1}\right]  \nonumber\\
	&\quad{}= \bbP \Big\{ (\bx_{i_1}, \bY_{i_2},\bx_{j_1},\bY_{j_2},\bw) \in \T(X,Y,X',Y',W) \text{ for some }(j_1,j_2)\in \tilde{A}_{(i_1,i_2)} \text{ and }j_2<i_2\nonumber\\
	&\hspace{28em}\Big|\, \bY_1,\ldots,\bY_{i_2-1}\Big\}\\
	&\quad{} \leq g_{i_1}(\bx_1,\ldots,\bx_{i_1})\label{assump1}\\
	&\quad{}\leq \exp(-n\delta(\eps)/2)\label{assump2}
\end{align}
where \eqref{assump1} and \eqref{assump2} follow directly from $g_{i_1}$ definition and our assumption for the $g_{i_1}$.
Thus, using Lemma \ref{LemA1} we get
\begin{align}
	\bbP\left(\frac{1}{N_2}\sum_{i_2=1}^{N_2}f_{(i_1,i_2)}(\bY_1,\ldots,\bY_{i_2})>\exp(-n\delta(\eps)/4)\right) \leq \exp(-\exp(n\sigma(\eps)))
\end{align}
where $\sigma(\eps)>0$.
If we sum over all $i_1$'s, we obtain
\begin{align}
	\sum_{i_1=1}^{N_1}\bbP\left(\frac{1}{N_2}\sum_{i_2=1}^{N_2}f_{(i_1,i_2)}(\bY_1,\ldots,\bY_{i_2})>\exp(-n\delta(\eps)/4)\right) \leq \exp(nR_1-\exp(n\sigma(\eps))),
\end{align}
that is this doubly exponential function still tends to zero as $n\to \infty$.
Therefore, with probability approaching $1$, for every $i_1$ that $g_{i_1}(\bx_1,\ldots,\bx_{i_1})\leq \exp(-n\delta(\eps)/2)$ we have
\begin{multline}
	\!\!\!\frac{1}{N_2}\Big|\!\left\{i_2: (\bx_{i_1}, \by_{i_2},\bx_{j_1},\by_{j_2},\bw) \!\in\! \T(X,Y,X',Y',W) \text{ for some }(j_1,j_2)\!\in\! \tilde{A}_{(i_1,i_2)} \text{ and }j_2\!<\!i_2\right\}\!\Big|\\
	\leq \exp(-n\delta(\eps)/4)\label{N2less}.
\end{multline}
In general, for all $i_1$ we have
\begin{align}
	&\frac{1}{N_1N_2}\Big|\Big\{(i_1,i_2): (\bx_{i_1}, \by_{i_2},\bx_{j_1},\by_{j_2},\bw) \in \T(X,Y,X',Y',W) \nonumber \\
	&\hspace{18em}\text{ for some }(j_1,j_2)\in \tilde{A}_{(i_1,i_2)} \text{ and }j_2<i_2\Big\}\Big|\label{thirdprob}\\
	& \quad{}\leq \frac{1}{N_1}\sum_{i_1=1}^{N_1}\frac{1}{N_2}\Big|\Big\{i_2: (\bx_{i_1}, \by_{i_2},\bx_{j_1},\by_{j_2},\bw) \in \T(X,Y,X',Y',W)\nonumber\\ 
	&\hspace{18em}\text{ for some }(j_1,j_2)\in \tilde{A}_{(i_1,i_2)} \text{ and }j_2<i_2\Big\}\Big|\\
	& \quad{} \leq \exp(-n\delta(\eps)/4)\!+\!\frac{1}{N_1}\!\!\!\!\!\sum_{i_1:g_{i_1}\leq \exp(-n\delta(\eps)/2)}\!\frac{1}{N_2}\Big|\Big\{i_2: (\bx_{i_1}, \by_{i_2},\bx_{j_1},\by_{j_2},\bw) \!\in\! \T(X,Y,X',Y',W)\label{fromg1}\nonumber\\
	&\hspace{18em}\text{ for some }(j_1,j_2)\in \tilde{A}_{(i_1,i_2)} \text{ and }j_2<i_2\Big\}\Big|\\
	& \quad{}\leq 2\exp(-n\delta(\eps)/4)\label{2_exp}
\end{align}
where \eqref{fromg1} and \eqref{2_exp} follow from \eqref{g1less} and \eqref{N2less}, respectively.

We may use this argument to upper bound the probability in \eqref{thirdprob} for the remain three cases $(j_1<i_1, j_2>i_2)$, $(j_1>i_1, j_2<i_2)$
and $(j_1>i_1, j_2>i_2)$ by defining different $A_{(i_1,i_2)}$'s, and conclude the same decreasing exponential function. Finally, we obtain
\begin{multline}
\frac{1}{N_1N_2}\big|\big\{(i_1,i_2):(\bx_{i_1},\by_{i_2},\bx_{j_1},\by_{j_2},\bw)\!\in\!\T(X,Y,X',Y',W)\text{ for some }j_1\ne i_1,j_2\ne i_2\big\}\big|\\
\leq 8\exp\{-n\delta(\eps)/4\}\end{multline}
if we have $J_{X';Y';W}(R_1,R_2)\le I(XY;X'Y'W)-2\delta(\eps)$.

In order to complete the proof, since for any fixed $\nu$ the cardinality of finite set $\scS^n$ is only increasingly exponentially in $n$, and the set $\calV^{5\times 5}$ is finite along with the doubly decreasing exponential probabilities in \eqref{eq:doubly_exp_c1} and \eqref{eq:doubly_exp_c2}, we derive that with probability approaching to $1$, all inequalities in \eqref{eq:two_codebooks_c0}, \eqref{eq:two_codebooks_c1}, \eqref{eq:two_codebooks_c2} and \eqref{eq:two_codebooks_c3} hold simultaneously for sufficiently large $n$. Since these inequalities hold for every element in the finite sets $\scS^n$ and $\calV^{5\times 5}$, then for any vector $\bw, \bx, \by$ and any given covariance matrix $\cov(X,Y,X',Y',W)$ (with $\|\bx\|^2= n\theta,\|\by\|^2= n\bar{\theta}, \|\bw\|^2\leq n\Lambda$) which is not in its corresponding $\nu$-dense subset, there exists a point in the corresponding $\nu$-dense subset that is close enough to it (in its $\nu$ distance neighborhood). Now, by using the continuity properties of all sets, we may conclude that \eqref{eq:two_codebooks_c0}, \eqref{eq:two_codebooks_c1}, \eqref{eq:two_codebooks_c2} and \eqref{eq:two_codebooks_c3} hold also for any point which is not in the $\nu$-dense subset.

\bibliographystyle{IEEEtran}

\end{document}

%% file: header.tex
\usepackage{graphicx,epsf,psfrag}
\usepackage{amsmath,amssymb}
\usepackage{amsfonts}
\usepackage{mathrsfs}
\usepackage{etoolbox}
\usepackage{cite}
\usepackage{bbm}
\usepackage{latexsym}

\newcommand{\beq}{\begin{equation}}
\newcommand{\eeq}{\end{equation}}
\newcommand{\be}{\begin{equation}}
\newcommand{\ee}{\end{equation}}
\newcommand{\eps}{\epsilon}

\newcommand{\bi}{\begin{itemize}}
\newcommand{\ei}{\end{itemize}}

\newcommand{\calE}{\mathcal{E}}

\newcommand{\calN}{\mathcal{N}}

\newcommand{\calT}{\mathcal{T}}
\newcommand{\calU}{\mathcal{U}}
\newcommand{\calV}{\mathcal{V}}

\newcommand{\bI}{\mathbf{I}}

\newcommand{\bV}{\mathbf{V}}
\newcommand{\bw}{\mathbf{w}}
\newcommand{\bW}{\mathbf{W}}
\newcommand{\bx}{\mathbf{x}}
\newcommand{\bX}{\mathbf{X}}
\newcommand{\by}{\mathbf{y}}
\newcommand{\bY}{\mathbf{Y}}
\newcommand{\bz}{\mathbf{z}}
\newcommand{\bZ}{\mathbf{Z}}

\newcommand{\bbE}{\mathbb{E}}

\newcommand{\bbP}{\mathbb{P}}

\newcommand{\bbR}{\mathbb{R}}

\newcommand{\scS}{\mathscr{S}}

\DeclareMathAlphabet{\mathbsf}{OT1}{cmss}{bx}{n}
\DeclareMathAlphabet{\mathssf}{OT1}{cmss}{m}{sl}

\DeclareSymbolFont{bsfletters}{OT1}{cmss}{bx}{n}  
\DeclareSymbolFont{ssfletters}{OT1}{cmss}{m}{n}
\DeclareMathSymbol{\bsfGamma}{0}{bsfletters}{'000}
\DeclareMathSymbol{\ssfGamma}{0}{ssfletters}{'000}
\DeclareMathSymbol{\bsfDelta}{0}{bsfletters}{'001}
\DeclareMathSymbol{\ssfDelta}{0}{ssfletters}{'001}
\DeclareMathSymbol{\bsfTheta}{0}{bsfletters}{'002}
\DeclareMathSymbol{\ssfTheta}{0}{ssfletters}{'002}
\DeclareMathSymbol{\bsfLambda}{0}{bsfletters}{'003}
\DeclareMathSymbol{\ssfLambda}{0}{ssfletters}{'003}
\DeclareMathSymbol{\bsfXi}{0}{bsfletters}{'004}
\DeclareMathSymbol{\ssfXi}{0}{ssfletters}{'004}
\DeclareMathSymbol{\bsfPi}{0}{bsfletters}{'005}
\DeclareMathSymbol{\ssfPi}{0}{ssfletters}{'005}
\DeclareMathSymbol{\bsfSigma}{0}{bsfletters}{'006}
\DeclareMathSymbol{\ssfSigma}{0}{ssfletters}{'006}
\DeclareMathSymbol{\bsfUpsilon}{0}{bsfletters}{'007}
\DeclareMathSymbol{\ssfUpsilon}{0}{ssfletters}{'007}
\DeclareMathSymbol{\bsfPhi}{0}{bsfletters}{'010}
\DeclareMathSymbol{\ssfPhi}{0}{ssfletters}{'010}
\DeclareMathSymbol{\bsfPsi}{0}{bsfletters}{'011}
\DeclareMathSymbol{\ssfPsi}{0}{ssfletters}{'011}
\DeclareMathSymbol{\bsfOmega}{0}{bsfletters}{'012}
\DeclareMathSymbol{\ssfOmega}{0}{ssfletters}{'012}

\DeclareMathOperator*{\argmin}{arg\,min}

\DeclareMathOperator{\var}{Var}

\DeclareMathOperator{\cov}{Cov}

\ifcsmacro{theorem}{}{
\newtheorem{theorem}{Theorem}
\newtheorem{lemma}[theorem]{Lemma}
\newtheorem{proposition}[theorem]{Proposition}

}

\newcommand{\qednew}{\nobreak \ifvmode \relax \else
      \ifdim\lastskip<1.5em \hskip-\lastskip
      \hskip1.5em plus0em minus0.5em \fi \nobreak
      \vrule height0.75em width0.5em depth0.25em\fi}